\documentclass[preprint,aps]{revtex4}

\usepackage{graphicx}    
\usepackage{dcolumn}     
\usepackage{bm}          

\begin{document}
\preprint{}

\title{Theory of enhanced proximity effect by mid gap Andreev resonant 
state in diffusive normal metal / triplet superconductor junctions }

\author{Y. Tanaka$^{1,2}$, S. Kashiwaya$^3$ and T. Yokoyama$^{1}$   
}
%
\affiliation{
$^1$Department of Applied Physics,
Nagoya University, Nagoya, 464-8603, Japan \\
$^2$
CREST, Japan Science and Technology Corporation (JST) 
Nagoya, 464-8603, Japan \\
$^3$NeRI of National Institute of Advanced Industrial Science and
Technology (AIST), Tsukuba, 305-8568, Japan 
}
%
\date{\today}
\begin{abstract}
Enhanced proximity effect by mid gap Andreev resonant state (MARS) 
in a diffusive normal metal / insulator / 
triplet superconductor (DN/TS) junction is 
studied based on the Keldysh-Nambu quasiclassical
Green's function formalism. 
By choosing a $p$-wave superconductor as a typical example of 
TS, conductance of the junction and the spatial variation of
quasiparticle local density of states (LDOS) in DN are calculated
as the function of the magnitudes of the resistance $R_{d}$, 
Thouless energy in DN and the transparency of the insulating barrier.
The resulting conductance spectrum has a 
zero bias conductance peak (ZBCP) and LDOS has a zero energy peak (ZEP) 
except for $\alpha = \pi/2$ $(0 \leq \alpha \leq \pi/2)$, 
where $\alpha$ 
denotes the angle between the lobe direction of the $p$-wave 
pair potential to the normal to the interface. 
The widths of the ZBCP and ZEP 
are reduced with the increase of $R_{d}$ while their heights are 
drastically enhanced. 
These peaks are revealed to be suppressed by applying a magnetic field.  
When the magnitude of $R_{d}/R_{0}$ is 
sufficiently large, 
the total  zero  voltage 
resistance of the  junction is almost independent of the 
$R_{d}$ for $\alpha \neq \pi/2$. 
The extreme case is $\alpha=0$, where total zero voltage
resistance is always $R_{0}/2$. 
We also studied the charge transport in 
$p_{x}+ip_{y}$-wave junctions, where only the quasiparticle 
with perpendicular injection feel MARS. 
Even in this case, the resulting LDOS in DN has a ZEP. 
Thus, the existence of the ZEP in LDOS of DN region is a 
remarkable feature for DN/TS junctions which have never been expected 
for DN/ singlet superconductor junctions where MARS and proximity effect compete with each other. 
Based on these results, a crucial test to identify triplet pairing 
superconductors 
based on tunneling experiments is proposed.
\end{abstract}
\maketitle

\section{Introduction}

Nowadays, enormous numbers of  unconventional 
superconductors have been  discovered where 
pair potential changes sign on the Fermi surface \cite{Ueda,Maeno,Scalapino}.
It is known that reflecting on the 
sign change of the pair potential \cite{Harlingen,Tsuei,Sigrist,Kashi00,Lof}, 
tunneling spectroscopy of unconventional superconductor 
is essentially phase sensitive \cite{Kashi00,TK95}. 
The most dramatic effect is the appearance  of 
zero bias conductance peak (ZBCP) 
in normal metal / insulator / unconventional superconductor junctions 
due to the formation of  mid gap Andreev resonant states  (MARS) 
\cite{Buch,96}. 
The ZBCP are observed in the 
actual many experiments 
\cite{Kashi00,e1,e2,e3,e4,e5,e6,e7,e8,e9,e10,e11,e12,e13,e14,e15}.

The MARS is a  unique resonant state 
expected for the interface of unconventional superconductor 
assisted by the Andreev reflection \cite{Buch,TK95,Andreev}. 
The origin of this MARS is due to the anomalous interference 
effect of quasiparticles 
at the interface where 
injected and reflected quasiparticles  
from the unconventional superconductor side 
feel different sign of the pair potentials \cite{Kashi00,TK95}. 
It is well known that 
MARS influences several charge transport properties 
\cite{T1,T2,T3,T4,T5,T6,T7,T8,TKJ,J0,J1,J2,J3,J4,J5,J6,J7,J8,J9}. 
Since MARS is expected both for triplet and singlet 
superconductor junctions \cite{tr1,tr2}, 
it is a  challenging issue 
to present a new idea to discriminate  triplet superconducting states  
\cite{tr1,tr2,tr3,tr4,tr5,tr6} from 
singlet ones through tunneling spectroscopy 
via MARS \cite{pwave2004}. 
\par
Although there are several studies about unconventional superconductor 
junctions, 
\cite{or1,or2,or3,dop1,dop2,dop3,dop4,dop5,f1,f2,f3,f4,f5,f6,f7,f8,f9,o1,o2,o3,o4,o5,o6,o7,o8,o9} 
almost all of them are restricted to  
ballistic regime. 
If we take into account impurity scattering, 
we can expect various interesting features 
even for conventional superconductor junctions
\cite{Beenakker1,Hekking,Giazotto,Klapwijk,Kastalsky,Nguyen,Wees,Nitta,Bakker,
Xiong,Magnee,Kutch,Poirier}. 
As regards diffusive normal metal / insulator / 
conventional singlet $s$-wave superconductor(DN/CSS) junctions, 
there has been a remarkable progress in theories of proximity effect 
\cite{Beenakker1,Lambert,Takane,reflec,Lesovik,Larkin,Volkov,
KL,Nazarov1,Yip,Stoof,Reentrance,Golubov,Takayanagi,Seviour,Belzig,minigap,Lambert2,Bezuglyi,Nazarov2,Golubov2003}. 
However, proximity effect in  unconventional superconductors \cite{Ohashi}
has been yet to be clarified. 
Recently, we have developed a theory of 
charge transport in diffusive normal metal / insulator /
unconventional singlet superconductor 
(DN/USS) junctions \cite{PRL2003,PRB2004} 
extending Nazarov's theory of matrix current \cite{Nazarov1,Nazarov2}
within quasiclassical treatment \cite{Zaitsev,Larkin,Volkov,Usadel}. 
First, this theory was applied to $d$-wave superconductor 
junctions \cite{PRB2004}. 
Unfortunately, however, it is revealed that 
the proximity effect and the MARS compete with each other in DN/USS junctions.
Although the interface resistance  is reduced by the
MARS irrespective  of the magnitude of the
transparency at the interface, 
the resulting total resistance of the junction 
$R$ is always larger than $R_{0}/2 + R_{d}$, 
where $R_{d}$ and $R_{0}$ is the resistance in DN and Sharvin resistance at the interface, respectively.
This is because that the 
angular average of many channels at the DN/USS interface
destruct the phase coherence of the MARS and the proximity
effect (see Fig. 1. of ref. \cite{pwave2004}).
This destructive angular average is due to
the sign change of the pair potentials felt by quasiparticles with
injection angle $\phi$ and those with $-\phi$,
where the angle $\phi$ is measured from the direction normal to the
junction interface.
However, in diffusive normal metal / insulator / triplet superconductor
(DN/TS) junctions, 
we can escape from the above destructive average.
In the last paper, we have presented a 
theory which is available for DN/TS junctions \cite{pwave2004}. 
It is revealed that 
charge transport in diffusive normal metal / triplet superconductor
(DN/TS) junctions is significantly unusual. 
We can expect  enhanced proximity effect by the MARS.
The  total zero voltage resistance $R$ in the DN/TS junctions
is significantly  reduced by the enhanced proximity effect in the 
presence of the MARS. 
It is remarkable that when the resistance in DN, $R_{d}$, 
is sufficiently larger than the Sharvin resistance $R_{0}$,
$R$ is  given by $R=R_{0}/C_{-}$,@which can become much smaller than the preexisting@lower limit value of $R$, $i.e.$,
$R_{0}/2 +R_{d}$.
In the above, $C_{-}$ is a constant completely
independent of both $R_{d}$ and $R_{b}$, where $R_{b}$ denotes the
interface resistance in the normal state.
When all quasiparticles injected at the interface feel the
MARS, $R$ is reduced to be
$R=R_{0}/2$ irrespective of the  magnitude of $R_{d}$ and $R_{b}$.
At the same time,  local density of states (LDOS) 
in the DN region has zero energy peak (ZEP)
due to the penetration of the MARS into the DN region 
from the triplet superconductor (TS) side of the DN/TS interface. 
These outstanding features have never been expected
either in DN/CSS junctions or DN/USS junctions. However, the reference
\cite{pwave2004} does not contain the necessary technical
details of the matrix current derivation and the obtained 
results are limited. 
In order to understand this remarkable enhanced  proximity effect by MARS 
peculiar to the DN/TS junctions much more in detail, 
one has to
evaluate the conductance spectrum and quasiparticle 
density of states in wide range of several parameters.\par
In this paper, we present a detailed derivation of the matrix current
in  (DN/TS) junctions.
We express compact formula of matrix current 
and  Keldysh-Nambu(KN) Green's functions
relevant to the actual boundary condition 
in comparison with those of DN/USS junctions. 
Here, we restrict our attention to triplet superconductors 
with $S_{z}=0$, 
where $S_{z}$ denotes the $z$ component of the total spin of a 
Cooper pair. 
In order to show up the anomalous charge transport in 
DN/TS junctions due to the coexistence of the MARS and the proximity effect, 
we present detailed
numerical calculations of the conductance spectra of DN/TS
junctions for $p$-wave
superconductors. We investigate the dependence of the spectra of 
the bias voltage conductance 
 on various parameters: the height of the barrier at the
interface, resistance $R_{d}$ in DN, the Thouless energy $E_{Th}$ in DN and
the angle  $\alpha $ between the normal to the interface and the lobe direction  of $p$-wave superconductor. 
$E_{Th}$ can be expressed by $E_{Th}=\hbar D /L^{2}$ with diffusion constant 
$D$ in DN and the length of DN. 
We  normalize  the voltage-dependent
conductance $\sigma_{S}(eV)$
by its value in the normal state, $\sigma_N$,
so that $\sigma_{T}(eV)=\sigma_{S}(eV)/\sigma_N$.
At the same time, we focus on the spatial dependence 
of LDOS in DN for DN/TS junctions. 
We also studied about several related physical quantities which identify the 
anomalous charge transport. 
Our main results are as follows: \par

\noindent 1. The ZBCP is always seen in the shape of $\sigma_{T}(eV)$ for 
$\alpha \neq \pi/2$ with $0 \leq \alpha \leq \pi/2$. 
The magnitude of $\sigma_{T}(0)$ is drastically enhanced 
with the increase of $R_{d}/R_{b}$. 
The half width of the ZBCP, which is proportional to 
$\exp(-C_{c}R_{d}/R_{0})$ for $C_{c}$,  is almost  independent of the transparency of the DN/TS interface and $E_{Th}$. 
On the other hand,  $\sigma_{T}(eV)$ for 
DN/USS junctions with $d_{xy}$-wave superconductor shows a very different 
behavior. The magnitude of $\sigma_{T}(eV)$ is reduced with the increase of 
$R_{d}/R_{b}$ due to the absence of the proximity effect. 
\par

\noindent 2. The LDOS in DN has a  ZEP except for the case with 
$\alpha=\pi/2$ where MARS is absent. The existence of the ZEP in DN means that 
the penetration of the MARS into DN. 
The height of the ZEP is significantly 
enhanced  with the increase of $R_{d}/R_{0}$. 
The half width of the ZEP is proportional to 
$\exp(-C_{\rho}R_{d}/R_{0})$ where 
$C_{\rho}$ is a constant almost independent of $Z$. 
\par

\noindent 3. We can express the LDOS in the DN region 
using the proximity parameter $\theta$ as ${\rm Real}[\cos \theta]$. 
If we denote the $\theta$ at DN/TS interface as $\theta_{0}$, 
$\theta_{0}=2iR_{d} \cos\alpha/R_{0}$ is satisfied at $\varepsilon=0$ 
where  quasiparticle energy  $\varepsilon$  is measured from the Fermi energy. 
Since  $\theta_{0}$ is a pure imaginary number for 
$\varepsilon=0$, we can expect a ZEP in LDOS. 
This unique feature  has never been expected for 
DN/USS of DN/CSS cases, where $\theta_{0}$ at $\varepsilon=0$ is always a real 
number.  \par

\noindent 4. 
The  total zero voltage resistance $R$ in the DN/TS junctions 
is significantly  reduced by the enhanced proximity effect in the 
presence of the MARS. 
It is remarkable that when $R_{d}$ is sufficiently
larger than the Sharvin resistance $R_{0}$,
$R$ is reduced to be $R=R_{0}/(2 \cos \alpha)$,
which can become much smaller than the preexisting
lower limit value of $R$, $i.e.$,
$R_{0}/2 +R_{d}$.
For low transparent junctions, 
$R$ is also reduced to be $R=R_{0}/(2 \cos \alpha)$. 
When all quasiparticles injected at the interface feel the
MARS, $R$ is reduced to be
$R=R_{0}/2$ irrespective of the  magnitude of $R_{d}$ and $R_{b}$.
This drastic situation is realized for $\alpha=0$.  \par
\noindent 6. 
The sharp ZBCP or ZEP in LDOS 
due to the enhanced proximity effect is 
sensitive to the applied magnetic field $H$. 
The height of ZBCP is significantly reduced by $H$. 
The threshold value of the magnetic field is 
$H_{Th} \simeq 8\hbar/(eS_{DN})$ where $S_{DN}$ denotes the 
magnitude of the area of DN region. \par
\noindent 7. 
As a candidate of TS (triplet superconductor), 
we also choose $p_{x}+ip_{y}$-wave superconductor. 
This superconducting state is so called chiral superconducting state 
where broken time reversal symmetry state  (BTRSS) is realized. 
Although only quasiparticles 
with perpendicular injection feel MARS, both LDOS and $\sigma_{T}(eV)$ 
has ZEP and ZBCP due to the enhanced proximity effect by MARS. \par
It should be remarked that 
these remarkable features have never been expected
either in DN/CSS or DN/USS junctions. 
The structure of the paper is as follows. We formulate the model
in use in section 2. We also present there the detailed derivation
of the matrix current and end up with the expression for the
normalized conductance. We focus on $p$-wave superconductor
junctions in section 3 and
evaluate $\sigma _{T}(eV)$, $\rho(\varepsilon)$, zero voltage resistance 
 and the measure of the proximity effect $%
\theta_{0}$ for various cases. The last part of this section, 
we discuss $p_{x}+ip_{y}$-wave junctions. 
We summarize the results in section 4.

\section{Formulation}

\begin{figure}[bh]
\begin{center}
\scalebox{0.7}{
\includegraphics[width=15.0cm,clip]{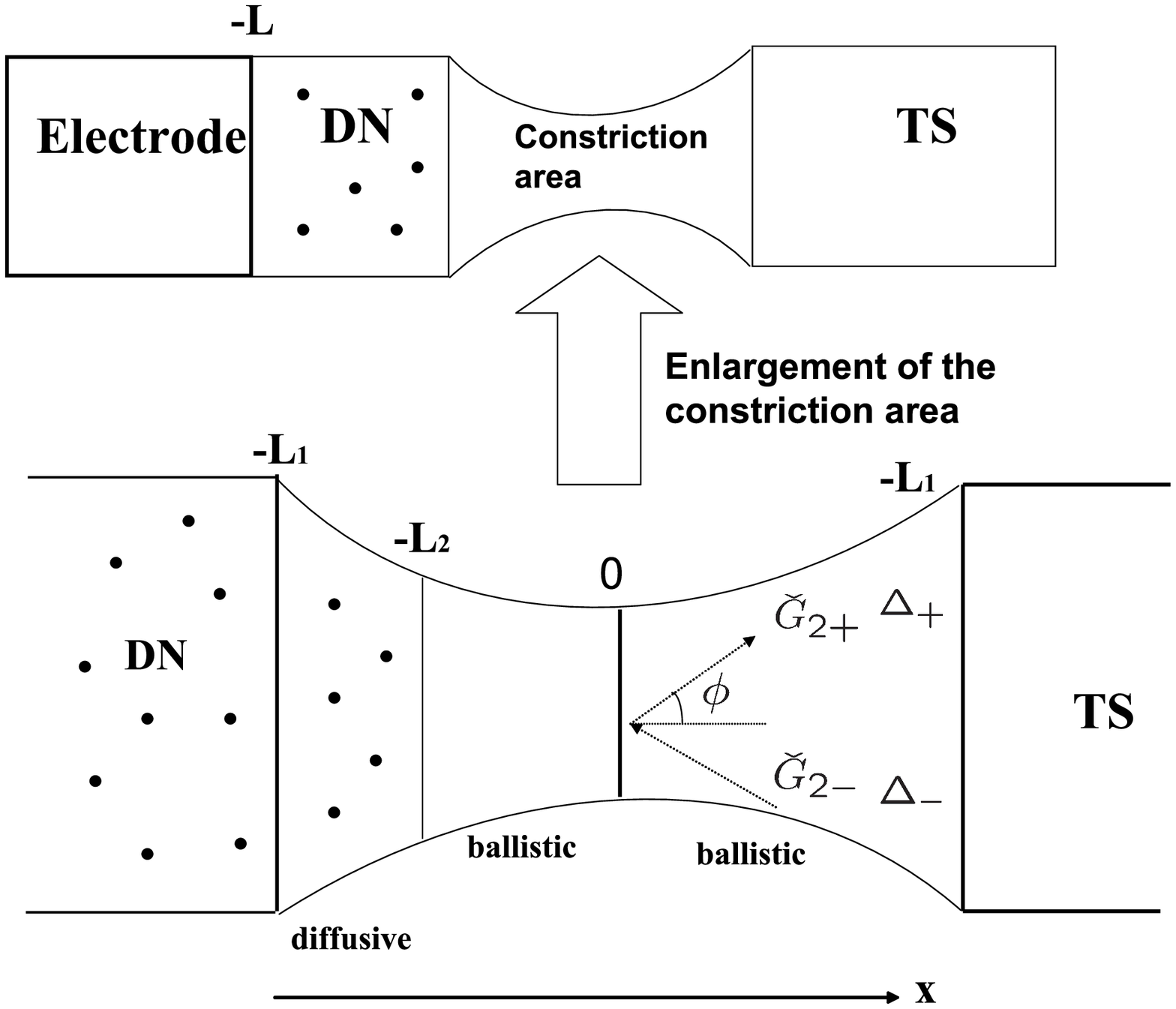}}
\end{center}
\caption{ Schematic illustration of the model 
of constriction area.  }
\label{fig:0}
\end{figure}

In this section we introduce the model and the formalism. We consider a
junction consisting of normal and superconducting reservoirs connected by a
quasi-one-dimensional diffusive conductor (DN) with a length $L$ much larger
than the mean free path as in our previous paper. 
The interface between the DN conductor and the TS
(triplet superconductor)  has a resistance $R_{b}$ while the
DN/N interface has zero resistance. A schematic illustration of 
the model we use are shown in Fig. \ref{fig:0}. 
The positions of the DN/N interface and
the DN/TS interface are denoted as $x=-L$ and $x=0$, respectively. According
to the circuit theory \cite{Nazarov2}, the constriction area ($-L_{1}<x<L_{1}
$) between DN and TS is considered as composed of the diffusive
isotropization zone $(-L_{1}<x<-L_{2})$, the left side ballistic zone $%
(-L_{2}<x<0)$, the right side ballistic zone $(0<x<L_{1})$ and the
scattering zone $(x=0)$. The scattering zone is modeled as an insulating
delta-function barrier with the transparency $T=T(\phi)=4\cos^{2}\phi /(4\cos
^{2}\phi +Z^{2})$, where $Z$ is a dimensionless constant, $\phi $ is the
injection angle measured from the interface normal to the junction. 
%
Here, we express insulating barrier as a delta function model 
$H_{b}\delta (x)$, 
where $Z$ is given by $Z=2mH_{b}/(\hbar ^{2}k_{F})$ with Fermi momentum $k_{F}$
and effective mass $m$.
In order to clarify charge transport in DN/TS junctions, we must
obtain Keldysh-Nambu (KN) Green's function, which has indices of
transport channels and the direction of motion along $x$ axis
taking into account the proper boundary conditions. 
We restrict our attention to triplet  superconductors
with $S_{z}=0$ that preserves time reversal symmetry.
$S_{z}$ denotes the
$z$ component of the total spin of a Cooper pair.
It is by no means easy
to formulate a charge transport of DN/TS junctions 
since  the quasiparticle
Green's function has no angular dependence by the impurity scattering in 
the DN. 
However, as shown in our previous paper \cite{PRL2003,PRB2004},
if we concentrate on the matrix currents
\cite{Nazarov1,Nazarov2} via the TS to or from the DN,
we can make a boundary condition of the KN Green's function. 
The sizes of the ballistic zone in the DN
and the scattering zone in the current flow direction are much
shorter than the coherence length \cite{PRL2003,PRB2004,Nazarov2}.
Since we assume the flat interface, the momentum parallel to the
interface is conserved.
Thus it is possible to construct a matrix current \cite{Nazarov1,Nazarov2}
based on the asymptotic
Green's function in TS.

In this section, we will show how to derive the matrix current in DN/TS
junctions. Then we will derive the retarded and Keldysh components of the
matrix current. Finally, we will show how to calculate the 
conductance of the junctions.

\subsection{Usadel equation in DN}
In order to clarify the charge transport in DN/TS junctions, 
we first concentrate on the KN Green's function in DN. 
We define KN Green's function in DN as 
\begin{equation}
\\
\check{G}_{N}(x) = \left(
\begin{array}{cc}
\hat{R}_{N}(x) & \hat{K}_{N}(x) \\
0 & \hat{A}_{N}(x)%
\end{array}
\right). 
\end{equation}
Here, we are using quasiclassical approximation. 
Then,  we can express $\hat{R}_{N}(x)$ as 
\begin{equation}
\hat{R}_{N}(x)=
\cos \psi \sin \theta \hat{\tau}_{1}
+ \sin \psi \sin \theta \hat{\tau}_{2}
+ \cos \theta \hat{\tau}_{3}, 
\end{equation}
with Pauli matrix in the electron hole space, 
$\hat{\tau}_{1}$, $\hat{\tau}_{2}$, and $\hat{\tau}_{3}$. 
Since $\hat{R}_{N}(x)$ obeys Usadel equation, 
following equations are satisfied, 
\begin{equation}
\hbar D[\frac{\partial^{2}}{\partial x^{2}}\theta 
- (\frac{\partial \psi}{\partial x})^{2}\cos \theta \sin \theta] 
+ 2i\varepsilon \sin\theta =0,
\end{equation}
\[
\frac{\partial}{\partial x}
[\sin^{2}\theta (\frac{\partial \psi}{\partial x})]=0.
\]
In the present case, since no supercurrent is flowing in the junctions, 
$\frac{\partial \psi}{\partial x}=0$ is satisfied. 
We can choose $\psi=\psi_{0}$ where $\psi_{0}$ is a constant 
independent of $x$. 
The boundary condition of $\check{G}_{N}(x)$ at DN/TS interface 
is given by 
\begin{equation}
\frac{L}{R_{d}}[\check{G}_{N}(x)\frac{\partial \check{G}_{N}(x)}{\partial x}%
]_{\mid x=0_{-}}=-\frac{<\check{I}(\phi)>}{R_{b}}.  \label{Nazarov}
\label{boundary1}
\end{equation}%
\[
\check{I}(\phi)=2[\check{G}_{1},\check{B}]
\]%
\begin{equation}
\check{B}=(-T_{1}[\check{G}_{1},\check{H}_{-}^{-1}]+\check{H}_{-}^{-1}%
\check{H}_{+}-T_{1}^{2}\check{G}_{1}\check{H}_{-}^{-1}\check{H}_{+}%
\check{G}_{1})^{-1}(T_{1}(1-\check{H}_{-}^{-1})+T_{1}^{2}\check{G}_{1}%
\check{H}_{-}^{-1}\check{H}_{+})
\end{equation}%
with $\check{G}_{1}=\check{G}_{N}(x=0_{-})$, 
$\check{H}_{\pm}=(\check{G}_{2+} \pm \check{G}_{2-})/2$, 
and $T_{1}=T/(2-T + 2\sqrt{1 -T})$ 
where $\check{G}_{2\pm}$ is the asymptotic Green's function 
in TS as defined in our previous papers. 
In the above, 
the average over the various angles of injected particles at the interface
is defined as
\begin{equation}
<\check{I}(\phi )>=\int_{-\pi /2}^{\pi /2}d\phi \cos \phi \check{I}(\phi
)/\int_{-\pi /2}^{\pi /2}d\phi T(\phi )\cos \phi
\label{average}
\end{equation}
with $\check{I}(\phi )=\check{I}$ and $T(\phi )=T$. The resistance 
of the interface $R_{b}$ is given by
\begin{equation}
R_{b}=\frac{2R_{0}}{\int_{-\pi /2}^{\pi /2}d\phi T(\phi )\cos
\phi }
\end{equation}
with Sharvin resistance at the interface, $R_{0}$. 
In the above, $\check{G}_{1}$ and $\check{G}_{2\pm}$ can be given by 
\begin{equation}
\check{G}_{1} =\left(
\begin{array}{cc}
\hat{R}_{1} & \hat{K}_{1} \\
0 & \hat{A}_{1}%
\end{array}%
\right), \ \ \check{G}_{2\pm}=\left(
\begin{array}{cc}
\hat{R}_{2\pm} & \hat{K}_{2\pm} \\
0 & \hat{A}_{2\pm}%
\end{array}%
\right),
\end{equation}%
where the Keldysh component $\hat{K}_{1,2\pm}$ is given by $\hat{K}_{1(2\pm)}=\hat{%
R}_{1(2\pm)}\hat{f}_{1(2)}(0)-\hat{f}_{1(2)}(0)\hat{A}_{1(2\pm)}$ with the
retarded component $\hat{R}_{1,2\pm}$ and
the advanced component $\hat{A}_{1,2\pm}$
using distribution function $\hat{f}_{1(2)}(0)$. In the above, $\hat{R}_{2\pm}$
is expressed by
\[
\hat{R}_{2\pm}=(g_{\pm}\hat{\tau}_{3}+f_{\pm}\hat{\tau}_{2})
\]%
with $g_{\pm}=\varepsilon /\sqrt{\varepsilon ^{2}-\Delta _{\pm}^{2}}$, 
$f_{\pm}=\Delta _{\pm}/\sqrt{\Delta _{\pm}^{2}-\varepsilon ^{2}}$, 
and $\hat{A%
}_{2\pm}=-\hat{\tau}_{3}\hat{R}_{2\pm}^{\dagger }\hat{\tau}_{3}$ where 
$\varepsilon $ denotes the quasiparticle energy measured from the Fermi
energy. 
$\Delta_{+}$ $(\Delta_{-})$  is the effective pair 
potential felt by quasiparticles with an injection angle $\phi$ $(\pi-\phi)$ 
as shown in Fig. \ref{fig:0}. 
$\hat{f}_{2}(0)=f_{0S}(0)=\mathrm{{tanh}[\varepsilon /(2k_{B}T)]}$
in thermal equilibrium with temperature $T$. Here, we put the electrical
potential zero in the TS. We also denote $\check{H}_{+}$, $\check{H%
}_{-}$, $\check{B}$, $\check{I}=\check{I}(\phi)$ as follows,
\begin{eqnarray}
\check{H}_{+} =\left(
\begin{array}{cc}
\hat{R}_{p} & \hat{K}_{p} \\
0 & \hat{A}_{p}%
\end{array}%
\right) , \ \ \check{H}_{-} =\left(
\begin{array}{cc}
\hat{R}_{m} & \hat{K}_{m} \\
0 & \hat{A}_{m}%
\end{array}%
\right),
\end{eqnarray}

\[
\check{B}=\left(
\begin{array}{cc}
\hat{B}_{R} & \hat{B}_{K} \\
0 & \hat{B}_{A}%
\end{array}%
\right) ,\ \ \check{I}=\left(
\begin{array}{cc}
\hat{I}_{R} & \hat{I}_{K} \\
0 & \hat{I}_{A}%
\end{array}%
\right).
\]%

\subsection{Calculation of the retarded part of the matrix current}
First, we pay attention to the boundary condition of the retarded part of 
KN Green's function at DN/TS interface in order to 
determine the value of $\psi_{0}$. 
The left side of the boundary condition of eq. (\ref{boundary1})
can be expressed by 
\begin{equation}
\frac{L}{R_{d}}
\hat{R}_{N}(x)\frac{\partial}{\partial x} \hat{R}_{N}(x) \vert_{x=0}
=\frac{Li}{R_{d}}[-\sin \psi_{0}\hat{\tau}_{1} + \cos \psi_{0} \hat{\tau}_{2}]
(\frac{\partial \theta}{\partial x}) \vert_{x=0}
\end{equation}
In the above, the most remarkable point is that the $\hat{\tau}_{3}$ component 
is vanishing. 
In order to calculate the right side of eq. (\ref{boundary1}), we must 
study the parity of $\hat{I}_{R}$ as a function of $\phi$. 
In general, $\hat{I}_{R}$ can be expressed by using several spectral 
vectors, 
\[
\hat{I}_{R}=4iT_{1}(\bm{ d}_{R} \cdot \bm{ d}_{R})^{-1}
\{ -\frac{1}{2}(1 + T_{1}^{2})({\bm s}_{2+}-{\bm s}_{2-})^{2}
[{\bm s}_{1}\times({\bm s}_{2+} + {\bm s}_{2-})]\cdot \hat{\bm{\tau}}
\label{spectral}
\]
\[
+2T_{1}{\bm s}_{1}\cdot({\bm s}_{2+} \times {\bm s}_{2-})
[{\bm s}_{1} \times ({\bm s}_{2+} \times {\bm s}_{2-})] \cdot \hat{\bm{\tau}}
\]
\[
+2T_{1}{\bm s}_{1}\cdot({\bm s}_{2+} - {\bm s}_{2-})
[{\bm s}_{1} \times ({\bm s}_{2+} - {\bm s}_{2-})] \cdot \hat{\bm{\tau}}
\]
\[
-i(1+T_{1}^{2})(1 -{\bm s}_{2+}\cdot{\bm s}_{2-})
[{\bm s}_{1} \times ({\bm s}_{2+} \times {\bm s}_{2-})]\cdot \hat{\bm{\tau}}
\]
\begin{equation}
+2iT_{1}(1 -{\bm s}_{2+}\cdot{\bm s}_{2-})
[{\bm s}_{1}\cdot({\bm s}_{2+}-{\bm s}_{2-}){\bm s}_{1} 
-({\bm s}_{2+} -{\bm s}_{2-})]\cdot \hat{\bm{\tau}}
\}
\end{equation}
\begin{equation}
\bm{d}_{R}=(1 + T_{1}^{2})(\bm{s}_{2+} \times \bm{s}_{2-})
-2T_{1}\bm{s}_{1}\times(\bm{s}_{2+}-\bm{s}_{2-})
-2T_{1}^{2}\bm{s}_{1}\cdot(\bm{s}_{2+}\times\bm{s}_{2-})\bm{s}_{1}
\end{equation}
with $\hat{R}_{1}={\bm s}_{1}\cdot\hat{\bm{\tau}}$ 
and  $\hat{R}_{2\pm}={\bm s}_{2\pm}\cdot\hat{\bm{\tau}}$.  

The spectral vectors ${\bm s}_{1}$ and ${\bm s}_{2\pm}$ are  given by 
\begin{equation}
{\bm s}_{1}
= 
\left(
\begin{array}{c}
\sin\theta_{0}\cos\psi_{0}  \\
\sin\theta_{0}\sin\psi_{0}  \\
\cos \theta_{0}
\end{array}%
\right); 
\bm{s}_{2\pm}=
\left(
\begin{array}{c}
0  \\
f_{\pm}(\phi)  \\
g_{\pm}(\phi)
\end{array}%
\right),
\end{equation}
where $\theta_{0}$ denotes the $\theta$ at $x=0_{-}$. 
We postulate that $\hat{\tau}_{3}$ component of $<\hat{I}_{R}>$ should be zero.
Here, we focus on the parity of $\hat{I}_{R}$ 
since there is an angular average over $\phi$ in the actual calculation. 
As a comparison, we also look at the case where TS (triplet superconductor) 
is substituted with USS (unconventional singlet superconductor). 
For DN/TS junctions, 
$g_{\pm}(-\phi)=g_{\mp}(\phi)$ and $f_{\pm}(-\phi)=-f_{\mp}(\phi)$ are 
satisfied, while for DN/USS junctions, 
$g_{\pm}(-\phi)=g_{\mp}(\phi)$ and $f_{\pm}(-\phi)= f_{\mp}(\phi)$ are 
satisfied.
$({\bm d}_{R} \cdot {\bm d}_{R})^{-1}$ can be given by 
\begin{equation}
({\bm d}_{R} \cdot {\bm d}_{R})^{-1}
=
\left\{
\begin{array}{cc}
d^{(t)}_{e}(\phi) + d^{(t)}_{o}(\phi)\sin\psi_{0} & {\rm DN/TS} 
\\
d^{(s)}_{e}(\phi)  & {\rm DN/USS }
\end{array}%
\right.
\end{equation}
for DN/TS and DN/USS junctions, 
where  $d^{(s)}_{e}(\phi)$  and $d^{(t)}_{e}(\phi)$ 
are  even  functions with $\phi$, while 
$d^{(t)}_{o}(\phi)$ 
is an  odd function with $\phi$. 
We apply similar discussions for other terms in Eq. (\ref{spectral}). 
For the convenience, we define 
\begin{equation}
{\bm s}_{z}=
\left(
\begin{array}{c}
0
\\
0
\\
1
\end{array}
\right)
\end{equation}
After simple manipulations, we can show the following relations 
\begin{equation}
 -\frac{1}{2}(1 + T_{1}^{2})({\bm s}_{2+}-{\bm s}_{2-})^{2}
[{\bm s}_{1}\times({\bm s}_{2+} + {\bm s}_{2-})]\cdot \bm{s}_{z}
\end{equation}
\[
=
\left\{
\begin{array}{cc}
F^{(t)}_{1o}(\phi)\cos\psi_{0} & {\rm DN/TS }\\ 
F^{(s)}_{1e}(\phi)\cos\psi_{0} & {\rm DN/USS }
\end{array}
\right.
\]

\begin{equation}
2T_{1}{\bm s}_{1}\cdot({\bm s}_{2+} \times {\bm s}_{2-})
[{\bm s}_{1} \times ({\bm s}_{2+} \times {\bm s}_{2-})] \cdot \bm{s}_{z}
\end{equation}

\[
=\left\{
\begin{array}{cc}
F^{(t)}_{2e}(\phi)\cos\psi_{0}\sin\psi_{0} & {\rm DN/TS} \\ 
F^{(s)}_{2e}(\phi)\cos\psi_{0}\sin\psi_{0} & {\rm DN/USS}
\end{array}
\right.
\]

\begin{equation}
2T_{1}{\bm s}_{1}\cdot({\bm s}_{2+} - {\bm s}_{2-})
[{\bm s}_{1} \times ({\bm s}_{2+} - {\bm s}_{2-})] \cdot \bm{s}_{z}
\end{equation}
\[
=\left\{
\begin{array}{cc}
(F^{(t)}_{3e}(\phi)\sin\psi_{0}  +F^{(t)}_{3o}) \cos\psi_{0} & {\rm DN/TS} \\ 
F^{(s)}_{3e}(\phi)\cos\psi_{0} & {\rm DN/USS}
\end{array}
\right.
\]

\begin{equation}
-i(1+T_{1}^{2})(1 -{\bm s}_{2+}\cdot{\bm s}_{2-})
[{\bm s}_{1} \times ({\bm s}_{2+} \times {\bm s}_{2-})]\cdot \bm{s}_{z}
\end{equation}
\[
=\left\{
\begin{array}{cc}
F^{(t)}_{4e}(\phi)\sin\psi_{0} & {\rm DN/TS} \\ 
F^{(s)}_{4o}(\phi)\sin\psi_{0} & {\rm DN/USS}
\end{array}
\right.
\]

\begin{equation}
2iT_{1}(1 -{\bm s}_{2+}\cdot{\bm s}_{2-})
[{\bm s}_{1}\cdot({\bm s}_{2+}-{\bm s}_{2-}){\bm s}_{1} 
-({\bm s}_{2+} -{\bm s}_{2-})]\cdot {\bm s}_{z}
\end{equation}

\[
=\left\{
\begin{array}{cc}
F^{(t)}_{5e}(\phi)\sin\psi_{0} + F^{(t)}_{5o}(\phi) & {\rm DN/TS} \\ 
F^{(s)}_{5o}(\phi)\sin\psi_{0} & {\rm DN/USS}
\end{array}
\right.
\]
In the above, 
$F^{(t)}_{2e}(\phi)$, $F^{(t)}_{3e}(\phi)$, $F^{(t)}_{4e}(\phi)$, 
$F^{(t)}_{5e}(\phi)$, 
$F^{(s)}_{1e}(\phi)$, $F^{(s)}_{2e}(\phi)$ and $F^{(s)}_{3e}(\phi)$, 
are even functions of $\phi$, 
while 
$F^{(t)}_{1o}(\phi)$, $F^{(t)}_{3o}(\phi)$,   
$F^{(s)}_{4o}(\phi)$, $F^{(t)}_{5o}(\phi)$, and  $F^{(s)}_{5o}(\phi)$
are odd functions of $\phi$, respectively. 
By postulating that the 
$\hat{\tau}_{3}$ component of $\hat{I}_{R}$ should vanish after the 
angular average over $\phi$, 
we can show that $\sin \psi_{0}=0$ for DN/TS junctions and 
$\cos \psi_{0}=0$ for DN/USS junctions. 

The resulting retarded part of the boundary condition 
is given by 
\begin{equation}
\frac{L}{R_{d}}
(\frac{\partial \theta}{\partial x})\vert_{x=0}
=\frac{<F>}{R_{b}}
\end{equation}
\begin{equation}
F=\frac{2T(f_{S}\cos\theta_{0}-g_{S}\sin\theta_{0})}
{2 -T + T(\cos\theta_{0}g_{S} + \sin\theta_{0}f_{S})}
\label{boundary2}
\end{equation}
with $g_{S}=(g_{+}+g_{-})/(1+g_{+}g_{-}+f_{+}f_{-})$,  
$f_{S}=i(f_{+}g_{-}-g_{+}f_{-})/(1+g_{+}g_{-}+f_{+}f_{-})$ 
for DN/TS junctions  and $g_{S}=(g_{+}+g_{-})/(1+g_{+}g_{-}+f_{+}f_{-})$, 
$f_{S}=(f_{+} + f_{-})/(1+g_{+}g_{-}+f_{+}f_{-})$ for 
DN/USS junctions, where $g_{\pm}=g_{\pm}(\phi)$ and 
$f_{\pm}=f_{\pm}(\phi)$. 
This is one of the central results of this paper. \par

\subsection{Calculation of the Keldysh part of the matrix current}

Next, we focus on the Keldysh component. We define $I_{b}$ 
as 
\begin{equation}
I_{b}=\frac{1}{4}{\rm Tr}[\hat{\tau}_{3}\hat{I}_{K}] \ \ 
\hat{I}_{K}
=2(\hat{R}_{1}\hat{B}_{K}+\hat{K}_{1}\hat{B}_{A}
-\hat{B}_{R}\hat{K}_{1}-\hat{B}_{K}\hat{A}_{1})
\end{equation}
with $\hat{K}_{1}=\hat{R}_{1}\hat{f}_{1}(0)-\hat{f}_{1}(0)\hat{A}_{1}$ $%
\hat{f}_{1}(0)=f_{0N}(0)+f_{3N}(0)\hat{\tau}_{3}$. 
$\hat{B}_{R}$ is given by 
\begin{equation}
\hat{B}_{R}
=
\left\{
\begin{array}{cc}
b^{(t)}_{1}\hat{\tau}_{1} + b^{(t)}_{2}\hat{\tau}_{2}+b^{(t)}_{3}\hat{\tau}_{3} & {\rm DN/TS} \\
b^{(s)}_{1}\hat{\tau}_{1}+b^{(s)}_{2}\hat{\tau}_{2}+b^{(s)}_{3}\hat{\tau}_{3} & {\rm DN/USS }
\end{array}
\right. 
\end{equation}
\[
b^{(t)}_{1} = -\frac{T_{1}(T_{1}\sin\theta_{0} + f_{S})}{\Lambda}, \ 
b^{(t)}_{2} = -\frac{T_{1}{\bar f}_{S}}{\Lambda}, \ 
b^{(t)}_{3} = -\frac{T_{1}(T_{1}\cos\theta_{0} + g_{S})}{\Lambda}, \ 
\]
\[
b^{(s)}_{1} = -\frac{T_{1}{\bar f}_{S}}{\Lambda}, \ 
b^{(s)}_{2} = -\frac{T_{1}(T_{1}\sin\theta_{0} + f_{S})}{\Lambda}, \ 
b^{(s)}_{3} = -\frac{T_{1}(T_{1}\cos\theta_{0} + g_{S})}{\Lambda}, \ 
\]
\[
\Lambda=(1+T_{1}^{2})+2T_{1}(g_{S}\cos\theta_{0} + f_{S} \sin\theta_{0})
\]
In the above $g_{S}$, $f_{S}$ and $\bar{f}_{S}$ 
are defined  by 
\begin{equation}
g_{S}=
\left\{
\begin{array}{cc}
(g_{+}+g_{-})/(1+g_{+}g_{-}+f_{+}f_{-}) & {\rm DN/TS} \\
(g_{+}+g_{-})/(1+g_{+}g_{-}+f_{+}f_{-}) & {\rm DN/USS} 
\end{array}
\right. 
\label{gs}
\end{equation}

\begin{equation}
f_{S}=
\left\{
\begin{array}{cc}
i(f_{+}g_{-}-f_{-}g_{+})/(1+g_{+}g_{-}+f_{+}f_{-}) & {\rm DN/TS} \\
(f_{+}+f_{-})/(1+g_{+}g_{-}+f_{+}f_{-}) & {\rm DN/USS} 
\end{array}
\right. 
\label{fs}
\end{equation}

\begin{equation}
\bar{f}_{S}=
\left\{
\begin{array}{cc}
(f_{+}+f_{-})/(1+g_{+}g_{-}+f_{+}f_{-}) & {\rm DN/TS} \\
i(f_{+}g_{-}- g_{+}f_{-})/(1+g_{+}g_{-}+f_{+}f_{-}) & {\rm DN/USS} 
\end{array}
\right. 
\label{bfs}
\end{equation}
We can calculate $\hat{B}_{A}$ similar to the case of $\hat{B}_{R}$. 
After simple manipulation, we can show that 
$\hat{B}_{A}=-\hat{\tau}_{3}\hat{B}_{R}^{\dagger}\hat{\tau}_{3}$
and the resulting $I_{b}$ is given by 

\begin{equation}
I_{b}=\mathrm{Trace}\{\hat{\tau}_{3}(\hat{R}_{1}\hat{B%
}_{K}+\hat{R}_{1}^{\dagger }\hat{B}_{K})
\end{equation}%
\[
-[\hat{\tau}_{3}(\hat{R}_{1}^{\dagger }\hat{B}_{R}^{\dagger }+\hat{B}_{R}%
\hat{R}_{1}+\hat{B}_{R}^{\dagger }\hat{R}_{1}+\hat{R}_{1}^{\dagger }\hat{B}%
_{R})]f_{0N}(0)-[(\hat{R}_{1}+\hat{R}_{1}^{\dagger })(\hat{B}_{R}+\hat{B}%
_{R}^{\dagger })]f_{3N}(0)\}/2,
\]%
After angular average over $\phi$ and the operation of the trace, 
it is easily shown that the second term which is proportional to 
$f_{0N}(0)$ does not contribute to $<I_{b}>$. 
It is necessary to obtain $\hat{B}_{K}$ which is given by
\begin{equation}
\hat{B}_{K}=\hat{D}_{R}^{-1}\hat{N}_{K}-\hat{D}_{R}^{-1}\hat{D}_{K}
\hat{B}_{A}
\end{equation}
with $\hat{B}_{A}=\hat{D}_{A}^{-1}\hat{N}_{A}$ and 
\[
\check{D}= -T_{1}[\check{G}_{1},\check{H}_{-}^{-1}] +
\check{H}_{-}^{-1}\check{H}_{+}
-T_{1}^{2}\check{G}_{1}\check{H}_{-}^{-1}\check{H}_{+}\check{G}_{1}, \ \
\check{D}=\left(
\begin{array}{cc}
\hat{D}_{R} & \hat{D}_{K} \\
0 & \hat{D}_{A}%
\end{array}%
\right),
\]
where $\hat{N}_{K}$ and $\hat{N}_{A}$ is the Keldysh and advanced part of
$\check{N}$ given by
\[
\check{N}=T_{1}-T_{1}\check{H}_{-}^{-1}+T_{1}^{2}\check{G}_{1}\check{H}_{-}^{-1}%
\check{H}_{+} \ \
\check{N}=\left(
\begin{array}{cc}
\hat{N}_{R} & \hat{N}_{K} \\
0 & \hat{N}_{A}%
\end{array}%
\right).
\]
We can express $\hat{N}_{K}$ and $\hat{D}_{K}$ as linear combinations of
distribution functions $f_{0S}(0)$, $f_{0N}(0)$, and $f_{3N}(0)$ as follows 
both for DN/TS and DN/USS junctions,

\begin{equation}
\hat{N}_{K}=
\left\{
\begin{array}{cc}
\hat{C}^{(t)}_{1}f_{0S}(0)+\hat{C}^{(t)}_{2}f_{0N}(0)+\hat{C}^{(t)}_{3e}f_{3N}(0), & {\rm DN/TS} \\
\hat{C}^{(s)}_{1o}f_{0S}(0)+\hat{C}^{(s)}_{2o}f_{0N}(0)+\hat{C}^{(s)}_{3o}f_{3N}(0), & {\rm DN/USS}
\end{array}
\right.
\end{equation}

\begin{equation}
\hat{D}_{K}=
\left\{
\begin{array}{cc}
\hat{C}^{(t)}_{4}f_{0S}(0)+\hat{C}^{(t)}_{5}f_{0N}(0)+\hat{C}^{(t)}_{6}f_{3N}(0), & {\rm DN/TS} \\
\hat{C}^{(s)}_{4o}f_{0S}(0)+\hat{C}^{(s)}_{5o}f_{0N}(0)+\hat{C}^{(s)}_{6o}f_{3N}(0), & {\rm DN/USS}
\end{array}
\right.
\end{equation}

\begin{equation}
\hat{D}_{R}^{-1}=
\left\{
\begin{array}{cc}
\hat{C}^{(t)}_{7e} +\hat{C}^{(t)}_{7o} & {\rm DN/TS }\\
\hat{C}^{(s)}_{7o} & {\rm DN/USS }
\end{array}
\right.
\end{equation}
with 
$\hat{C}^{(t)}_{1}=\hat{C}^{(t)}_{1e}+\hat{C}^{(t)}_{1o}$, 
$\hat{C}^{(t)}_{2}=\hat{C}^{(t)}_{2e}+\hat{C}^{(t)}_{2o}$, 
$\hat{C}^{(t)}_{4}=\hat{C}^{(t)}_{4e}+\hat{C}^{(t)}_{4o}$, 
$\hat{C}^{(t)}_{5}=\hat{C}^{(t)}_{5e}+\hat{C}^{(t)}_{5o}$, 
and 
$\hat{C}^{(t)}_{6}=\hat{C}^{(t)}_{6e}+\hat{C}^{(t)}_{6o}$, 
by $2\times 2$ matrix $\hat{C}^{(r)}_{jk}$ with $r=t,s$, 
$j=1,..,7$ and $k=e,o$.  
When the suffix $k$ is $e$ ($o$), $\hat{C}^{(r)}_{jk}$ is an even (odd) function of $\phi$. 
$\hat{C}^{(t)}_{1e}$, $\hat{C}^{(t)}_{2e}$, 
$\hat{C}^{(t)}_{4o}$, $\hat{C}^{(t)}_{5o}$,  $\hat{C}^{(t)}_{6e}$,
and $\hat{C}^{(t)}_{7o}$
are linear combinations of $\hat{1}$ and $\hat{\tau}_{2}$, while 
$\hat{C}^{(t)}_{1o}$, $\hat{C}^{(t)}_{2o}$,   $\hat{C}^{(t)}_{3e}$, 
$\hat{C}^{(t)}_{4e}$, $\hat{C}^{(t)}_{5e}$,  $\hat{C}^{(t)}_{6o}$
and $\hat{C}^{(t)}_{7e}$ 
are linear combinations of $\hat{\tau}_{1}$ and $\hat{\tau}_{3}$.
$\hat{1}$ is a unit matrix in the electron hole space. 
On the other hand, $\hat{C}^{(s)}_{1o}$, $\hat{C}^{(s)}_{2o}$ 
and $\hat{C}^{(s)}_{6o}$
are linear combinations of $\hat{\tau}_{2}$ and $\hat{\tau}_{3}$, while 
$\hat{C}^{(s)}_{3o}$, $\hat{C}^{(s)}_{4o}$,  $\hat{C}^{(s)}_{5o}$,   
and $\hat{C}^{(s)}_{7o}$ 
are linear combinations of $\hat{1}$ and $\hat{\tau}_{1}$.
Taking account of these facts, after angular average over $\phi$,  
$<I_{b}>$ can be given by 
\[
<I_{b}>
=<\mathrm{Trace}\{(\hat{R}_{1}+\hat{R}%
_{1}^{\dagger })[\hat{B}_{KE}\hat{\tau}_{3}
-(\hat{B}_{R}+\hat{B}_{R}^{\dagger })]f_{3N}(0)\}>/2,
\]%
\[
\hat{B}_{KE}=\hat{D}_{R}^{-1}[\hat{C}_{3}-\hat{C}_{6}\hat{D}_{A}^{-1}\hat{N}%
_{A}],
\]%
\[
\hat{C}_{3}=T_{1}^{2}(\hat{R}_{1}\hat{\tau}_{3}-\hat{\tau}_{3}\hat{A}_{1})%
\hat{A}_{m}^{-1}\hat{A}_{p},
\]%
\[
\hat{C}_{6}=T_{1}[-(\hat{R}_{1}\hat{\tau}_{3}-\hat{\tau}_{3}\hat{A}_{1})%
\hat{A}_{m}^{-1}+\hat{R}_{m}^{-1}(\hat{R}_{1}\hat{\tau}_{3}-\hat{\tau}_{3}%
\hat{A}_{1})
\]%
\begin{equation}
-T_{1}(\hat{R}_{1}\hat{\tau}_{3}-\hat{\tau}_{3}\hat{A}_{1})\hat{A}_{m}^{-1}%
\hat{A}_{p}\hat{A}_{1}-T_{1}\hat{R}_{1}\hat{R}_{m}^{-1}\hat{R}_{p}(\hat{R}%
_{1}\hat{\tau}_{3}-\hat{\tau}_{3}\hat{A}_{1})].
\end{equation}%
with $\hat{C}_{3}=\hat{C}^{(t)}_{3e}$ ($\hat{C}_{3}=\hat{C}^{(s)}_{3o}$) 
and $\hat{C}_{6}=\hat{C}^{(t)}_{6}$ ($\hat{C}_{6}=\hat{C}^{(s)}_{6o}$) 
for DN/TS  (DN/USS) junctions. \par
Since following equations are satisfied,
\begin{equation}
\hat{D}_{A}^{-1}\hat{N}_{A}=-\hat{\tau}_{3}\hat{B}_{R}^{\dagger }\hat{\tau}_{3},\ \
\hat{A}_{m(p)}=-\hat{\tau}_{3}\hat{R}_{m(p)}^{\dagger }\hat{\tau}_{3},
\end{equation}

\begin{equation}
\hat{D}_{R}^{-1}(T_{1}-T_{1}\hat{R}_{m}^{-1}+T_{1}^{2}\hat{R}_{1}\hat{R}%
_{m}^{-1}\hat{R}_{p})=\hat{B}_{R},
\end{equation}

\begin{equation}
\hat{B}_{R}
(1 + \hat{R}_{m}^{-1} + T_{1}\hat{R}_{1}\hat{R}_{p}\hat{R}_{m}^{-1})
=T_{1}\hat{R}_{m}^{-1}\hat{R}_{p}
\end{equation}
$<I_{b}>$ is given by
\[
<I_{b}>
=\frac{1}{2}<
{\rm Trace}
\{
-(\hat{R}_{1}+\hat{R}_{1}^{\dagger })\hat{B}%
_{R}(\hat{R}_{1}+\hat{R}_{1}^{\dagger })\hat{B}_{R}^{\dagger }-(\hat{R}_{1}+%
\hat{R}_{1}^{\dagger })(\hat{B}_{R}+\hat{B}_{R}^{\dagger })
\}
>f_{3N}(0)
\]
For the later convenience, 
we define $<I_{b0}>$ with 
$<I_{b0}>=<I_{b}>/f_{3N}(0)$. 
Then the final resulting 
expression of $<I_{b0}>$ is given by the following equation, 
\begin{equation}
<I_{b0}>=<
\frac{T}{2}\frac{C_{0}}
{\mid (2-T)+T(\cos \theta _{0}g_{S}
+\sin\theta _{0}f_{S}) \mid ^{2}} >
\label{ib0}
\end{equation}%
\[
C_{0}=T(1+\mid \cos \theta _{0}\mid ^{2}+\mid \sin \theta _{0}\mid ^{2})
\lbrack \mid g_{S}\mid ^{2}+\mid f_{S}\mid ^{2}+1
+\mid \bar{f}_{S}\mid ^{2}]
\]%
\[
+4(2-T)[\mathrm{Real}(g_{S})\mathrm{Real}(\cos\theta_{0}) 
+ \mathrm{Real}(f_{S})\mathrm{Real}(\sin\theta_{0}) ]
\]
\[
+ 4T[\mathrm{Imag}(\cos\theta_{0}\sin\theta_{0}^{*})
\mathrm{Imag}(f_{S}g_{S}^{*})]. 
\]
In the above, the definition of 
$g_{S}$, $f_{S}$ and $\bar{f}_{S}$ are given in 
eqs. (\ref{gs}),(\ref{fs}) and (\ref{bfs}). 
By choosing $\theta_{0}=0$, we can reproduce well known results 
in ballistic limit \cite{BTK,TK95,tr1}.

\subsection{Calculation of the conductance}
The electric current is expressed using $\check{G}_{N}(x)$ as
\begin{equation}
I_{el}=\frac{-L}{4eR_{d}}\int_{0}^{\infty }d\varepsilon \mathrm{Tr}[
\hat{\tau}_{3}(\check{G}_{N}(x)\frac{\partial \check{G}_{N}(x)}{\partial x})^{K}],
\end{equation}%
where $(\check{G}_{N}(x)\frac{\partial \check{G}_{N}(x)}{\partial x})^{K}$
denotes the Keldysh component of $(\check{G}_{N}(x)\frac{\partial \check{G}%
_{N}(x)}{\partial x})$. %
In the actual calculation, we introduce a parameter 
$\theta=\theta (x)$ which is a
measure of the proximity effect in DN as described in the previous 
subsections, 
where we denoted $\theta (0)=\theta_{0}$. 
Using $\theta (x)$, $\hat{R}_{N}(x)$ can
be denoted as
\begin{equation}
\hat{R}_{N}(x)=\hat{\tau}_{3}\cos \theta (x)+\hat{\tau}_{2}\sin \theta (x).
\end{equation}%
$\hat{A}_{N}(x)$ and $\hat{K}_{N}(x)$ satisfy the following equations, $\hat{%
A}_{N}(x)=-\hat{\tau}_{3}\hat{R}_{N}^{\dagger }(x)\hat{\tau}_{3}$, and $\hat{K}%
_{N}(x)=\hat{R}_{N}(x)\hat{f}_{1}(x)-\hat{f}_{1}(x)\hat{A}_{N}(x)$ with the
distribution function $\hat{f}_{1}(x)$ which is given by $\hat{f}%
_{1}(x)=f_{0N}(x)+\hat{\tau}_{3}f_{3N}(x)$. In the above, $f_{3N}(x)$ is the
relevant distribution function which determines the conductance of the
junction we are now concentrating on. From the retarded or advanced
component of the Usadel equation, the spatial dependence of $\theta (x)$ is
determined by the following equation
\begin{equation}
\hbar D\frac{\partial ^{2}}{\partial x^{2}}\theta (x)+2i\varepsilon \sin [\theta
(x)]=0,  \label{Usa1}
\end{equation}%
%
%
%
%
%
while for the Keldysh component we obtain
\begin{equation}
D\frac{\partial }{\partial x}[\frac{\partial f_{3N}(x)}{\partial x}\mathrm{%
cosh^{2}}\theta _{imag}(x)]=0.  \label{Usa2}
\end{equation}%
%
%
%
%
%
with $\theta _{imag}(x)=\mathrm{Imag}[\theta (x)]$. At $x=-L$, since DN is
attached to the normal electrode, $\theta (-L)$=0 and $f_{3N}(-L)=f_{t0}$ are 
satisfied with
\[
f_{t0}=\frac{1}{2}\{\tanh [(\varepsilon +eV)/(2k_{B}T)]-\tanh [(\varepsilon
-eV)/(2k_{B}T)]\},
\]%
where $V$ is the applied bias voltage.
As shown in our previous paper, 
from the Keldysh part of Eq.~(\ref{Nazarov}), we obtain
\begin{equation}
\frac{L}{R_{d}}(\frac{\partial f_{3N}}{\partial x})\mathrm{{\cosh ^{2}}}%
{\rm Imag}(\theta _{0})\mid _{x=0_{-}}=-\frac{<I_{b}>}{R_{b}}.  \label{b2}
\end{equation}
After a simple manipulation, we can obtain $f_{3N}(0_{-})$
\[
\displaystyle f_{3N}(0_{-})=
\frac{R_{b}f_{t0}}{R_{b}+\frac{R_{d}<I_{b0}>}{L}%
\int_{-L}^{0}\frac{dx}{\cosh ^{2}\theta _{imag}(x)}}. 
\]%
Since the electric current $I_{el}$ can be expressed via $\theta _{0}$ in
the following form%
\[
I_{el}=-\frac{L}{eR_{d}}\int_{0}^{\infty }(\frac{\partial f_{3N}}{\partial x}%
)\mid _{x=0_{-}}\cosh ^{2}[\mathrm{Imag}(\theta _{0})]d\varepsilon ,
\]%
%
we obtain the following final result for the current

\begin{equation}
I_{el}=\frac{1}{e}\int_{0}^{\infty }d\varepsilon \frac{f_{t0}}{\frac{R_{b}}{
<I_{b0}>}
+\frac{R_{d}}{L}\int_{-L}^{0}\frac{dx}{\cosh ^{2}\theta _{imag}(x)}}%
.
\end{equation}%
Then the total resistance $R$ at zero temperature is given by
\begin{equation}
R=\frac{R_{b}}{<I_{b0}>}+\frac{R_{d}}{L}\int_{-L}^{0}\frac{dx}{\cosh
^{2}\theta _{imag}(x)}.
\label{resistance}
\end{equation}
In the following section, we will discuss the normalized
conductance $\sigma _{T}(eV)=\sigma _{S}(eV)/\sigma _{N}(eV)$ where $\sigma
_{S(N)}(eV)$ is the voltage-dependent conductance in the superconducting (normal)
state given by $\sigma _{S}(eV)=1/R$ and $\sigma _{N}(eV)=\sigma
_{N}=1/(R_{d}+R_{b})$, respectively.

It should be remarked that in the present  theory, $R_{d}/R_{b}$ can
be varied independently of $T$, $i.e.$ of $Z$, since we can change the
magnitude of the constriction area independently. In the other words, $%
R_{d}/R_{b}$ is no more proportional to $T_{av}(L/l)$, where $T_{av}$ is the
averaged transmissivity and $l$ is the mean free path in the diffusive
region, respectively. Based on this fact, we can choose $R_{d}/R_{b}$ and $Z$
as independent parameters. 

\section{Results}

In this section, we focus on the line shapes of the  conductance
of DN/TS junctions 
where $p$-wave symmetry is chosen as a pairing symmetry of 
triplet superconductor (TS). 
The pair potentials $\Delta_{\pm}$ are given by 
$\Delta_{\pm}=\pm\Delta_{0}\cos[(\phi \mp \alpha)]$ where $\alpha$ denotes the
angle between the normal to the interface and the lobe direction of the 
$p$-wave pair potential and $\Delta_{0}$ is the maximum amplitude of the pair
potential. 
In the above, $\phi$ denotes the injection angle of the
quasiparticle measured from the $x$-axis. 
In the following, we choose $0 \leq \alpha \leq \pi/2$. 
It is known that quasiparticles
with injection angle $\phi$ with $-\pi/2  + \alpha < \phi <
\pi/2 -\alpha $ feel the MARS at the interface 
and  induce ZBCP.

\subsection{Line shapes of the  conductance }

\begin{figure}[bh]
\begin{center}
\scalebox{0.4}{
\includegraphics[width=15.0cm,clip]{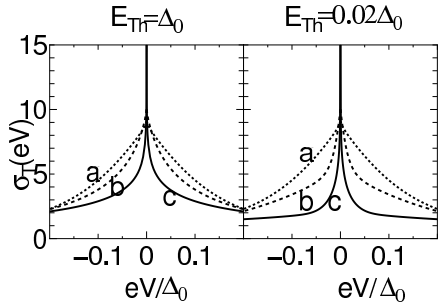}}
\end{center}
\caption{ Normalized  conductance $\sigma_{T}(eV)$ for Z=3, and $\protect\alpha=0$. $E_{Th}=\Delta_{0}$ for the left panel and $E_{Th}=0.02\Delta_{0}$ 
for the right panel, respectively. 
a, $R_{d}/R_{b}=0$; b, $R_{d}/R_{b}=0.1$; 
and c, $R_{d}/R_{b}=1$. }
\label{fig:1}
\end{figure}

\begin{figure}[bh]
\begin{center}
\scalebox{0.4}{
\includegraphics[width=15.0cm,clip]{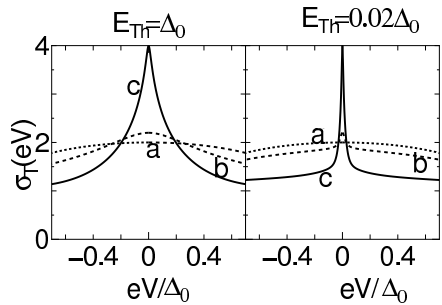}}
\end{center}
\caption{Normalized  conductance $\sigma_{T}(eV)$ for Z=0, and 
$\protect\alpha=0$. 
$E_{Th}=\Delta_{0}$ for the left panel and $E_{Th}=0.02\Delta_{0}$ 
for the right panel, respectively. 
a, $R_{d}/R_{b}=0$; b, $R_{d}/R_{b}=0.1$; 
and c, $R_{d}/R_{b}=1$. }
\label{fig:2}
\end{figure}

Let us first choose $\alpha=0$ where all quasiparticles feel 
MARS. For $alpha=0$,  we also call $p_{x}$-wave case in the following. 
In this case, at $eV=0$, 
the total resistance of the junction $R$ is $R_{0}/2$ and completely 
independent of $R_{d}$ as shown in our previous paper \cite{pwave2004}. 
Thus the resulting $\sigma_{T}(0)$ is 
$\sigma_{T}(0)=2(R_{d}+R_{b})/R_{0}$. 
In Figs. \ref{fig:1} and \ref{fig:2}, voltage dependent conductance 
$\sigma_{T}(eV)$ is plotted for low (Fig. \ref{fig:1})
and high transparent (Fig. \ref{fig:2}) interface. 
For $Z=3$ (Fig. \ref{fig:1}), we always expect ZBCP independent of the detailed value of $R_{d}/R_{b}$. 
With the increase of the magnitude of $R_{d}/R_{b}$, 
the width of the ZBCP becomes narrow 
(see curves $b$ and $c$ in the left panel),
while it's height is drastically enhanced. 
The width of the ZBCP is rather reduced by choosing small magnitude of $E_{Th}$(see curves $b$ and $c$ in the right panels of Fig. \ref{fig:1}). 

Similar plots for $Z=0$ is  shown in Fig. \ref{fig:2}. 
In this case, $R_{b}=R_{0}$ is satisfied due to the 
absence of the insulating barrier at the interface. 
For $R_{d}/R_{b}=0$, $\sigma_{T}(eV)$ has a broad ZBCP 
due to the absence of the normal reflection at the interface \cite{TK95}. 
With the increase of the magnitude of $R_{d}/R_{b}$, 
the width of the ZBCP is reduced  (see curves $b$ and $c$ both in the 
left and right panels), while its height is enhanced. 
As seen  from the right panel of Fig. \ref{fig:1}, 
the width of the ZBCP is reduced drastically 
with the decrease of the magnitude of  $E_{Th}$. 
%

\begin{figure}[tbh]
\begin{center}
\scalebox{0.4}{
\includegraphics[width=15.0cm,clip]{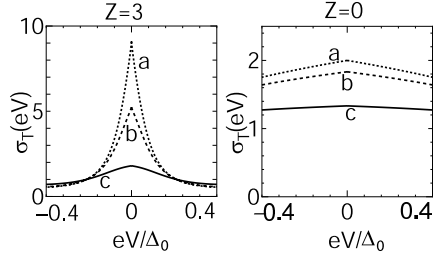}}
\end{center}
\caption{  Normalized  conductance $\sigma_{T}(eV)$ for DN/USS junctions 
with $d_{xy}$-wave superconductor. 
$Z=3$ for the left panel and $Z=0$ for the right panel. 
a, $R_{d}/R_{b}=0$; b,
$R_{d}/R_{b}=0.1$; and c, $R_{d}/R_{b}=1$. 
 }
\label{fig:3}
\end{figure}

In Fig. \ref{fig:3}, 
the corresponding plots for DN/USS (diffusive normal metal / insulator / 
unconventional singlet superconductor )junctions are shown 
with $d$-wave superconductor, where the misorientation angle between the 
normal to the interface and the crystal axis is 
chosen to be $\beta=\pi/4$. 
For $\beta=\pi/4$, we also call $d_{xy}$-wave case in the following. 
$\Delta_{\pm}$ is given by $\Delta_{\pm}= \pm \Delta_{0}\sin(2\theta)$. 
Although quasiparticles feel MARS independent 
of their injection angles, 
as shown in our previous papers \cite{PRB2004}, 
the proximity effect and MARS strongly compete with each other 
and the proximity effect is completely absent.
$\sigma_{T}(eV)$  is independent of  the value of $E_{Th}$. 
Both for $Z=0$ and $Z=3$, the magnitude of 
$\sigma_{T}(eV)$ is reduced monotonically with the 
increase of $R_{d}/R_{b}$. 
This feature is completely different from that in DN/TS junctions 
as shown in Figs. \ref{fig:1} and \ref{fig:2}. 

\begin{figure}[bh]
\begin{center}
\scalebox{0.4}{
\includegraphics[width=15.0cm,clip]{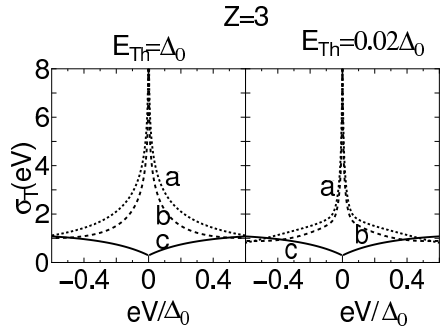}}
\end{center}
\caption{Normalized  conductance $\sigma_{T}(eV)$ for Z=3, and 
$R_{d}/R_{b}=1$. 
$E_{Th}=\Delta_{0}$ for the left panel and $E_{Th}=0.02\Delta_{0}$ 
for the right panel, respectively. 
a, $\alpha=0$; b, $\alpha=\pi/4$; 
and c, $\alpha=\pi/2$. }
\label{fig:4}
\end{figure}

Next, we look at  $\alpha$ dependence of $\sigma_{T}(eV)$ 
for DN/TS junctions. 
In this case, an injected quasiparticle with the injection angle 
$\phi$ with $-\pi/2 +  \alpha  < \phi < \pi/2 - \alpha$
feel the MARS. 
It is shown in our previous paper, that 
only the quasiparticle channel with this injection angle 
can contribute to the proximity effect. 
Only at $\alpha=\pi/2$, neither proximity effect nor MARS exist. 
As shown in  Fig. \ref{fig:4}, ZBCP is absent for $\pi/2$. 
In other cases, $\sigma_{T}(eV)$ has a ZBCP. 
%
%

\begin{figure}[tbh]
\begin{center}
\scalebox{0.4}{
\includegraphics[width=15.0cm,clip]{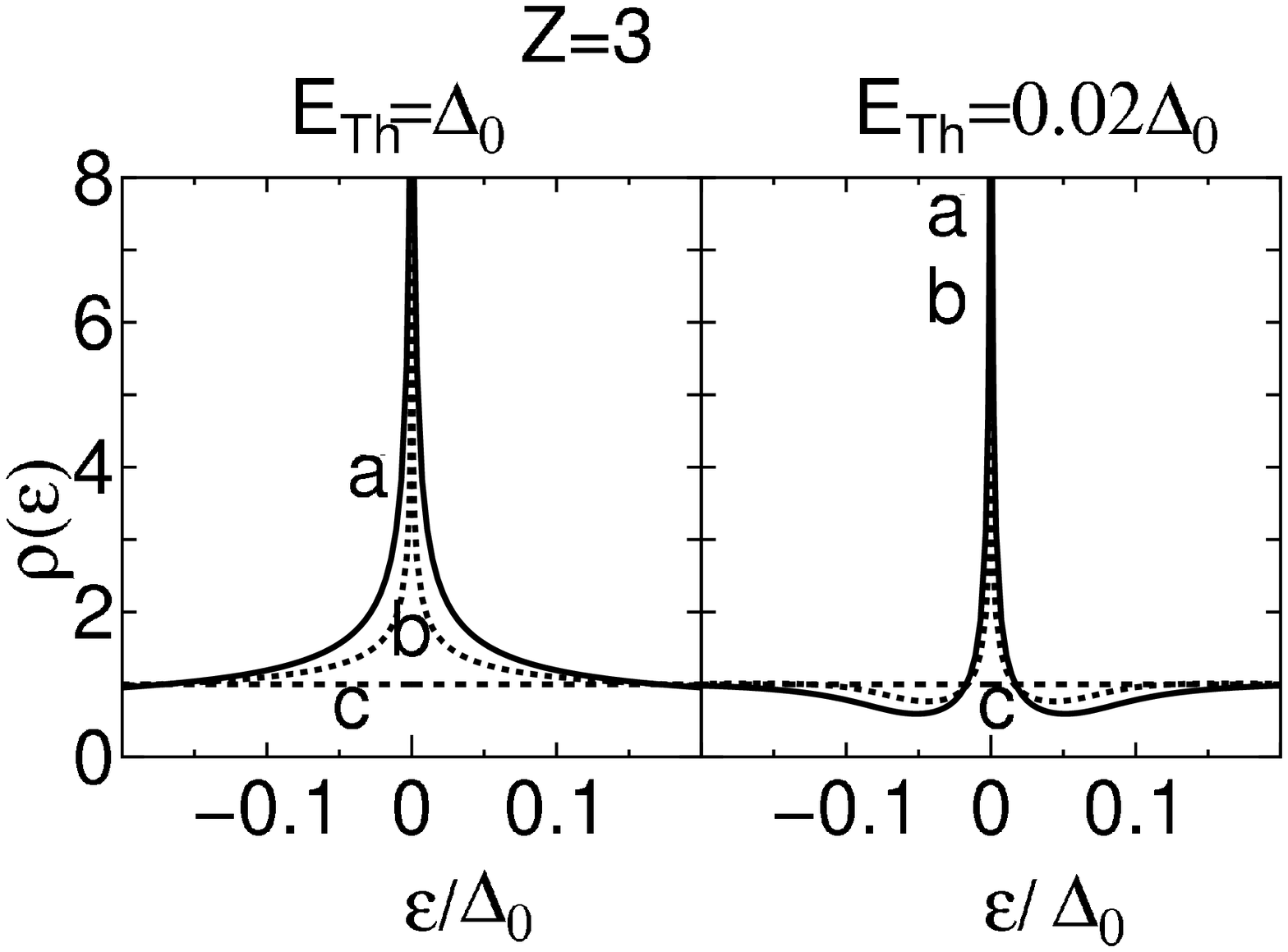}}
\end{center}
\caption{  Normalized  local density of states 
$\rho(\varepsilon)$ in DN for Z=3, $R_{d}/R_{b}=1$ and $\protect\alpha=0$. 
$E_{Th}=\Delta_{0}$ (left panel) and $E_{Th}=0.02\Delta_{0}$ 
(right panel). a, $x=-L/4$; b, $x= -L/2$; and c, $x=-L$. }
\label{fig:5}
\end{figure}

In order to clarify the proximity effect in DN, 
it is necessary to focus on the  
local density of states (LDOS) of the quasiparticles in DN region. 
In Fig. \ref{fig:5}, LDOS of DN/TS junctions is plotted 
for $\alpha=0$ with $E_{Th}=\Delta_{0}$ and $E_{Th}=0.02\Delta_{0}$. 

\begin{figure}[tbh]
\begin{center}
\scalebox{0.4}{
\includegraphics[width=18cm,clip]{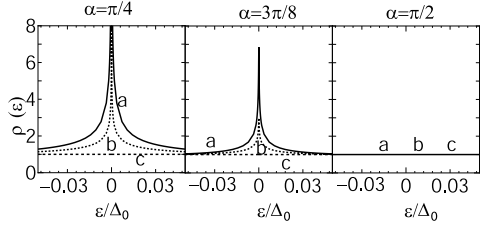}}
\end{center}
\caption{  Normalized  local density of states 
$\rho(\varepsilon)$ in DN for Z=3, $R_{d}/R_{b}=1$ and $E_{Th}=\Delta_{0}$. 
$\alpha=\pi/4$ (left panel), $\alpha=3\pi/8$ 
(middle panel), and $\alpha=\pi/2$ (right panel). 
a, $x=-L/4$; b, $x= -L/2$; and c, $x=-L$. }
\label{fig:6}
\end{figure}

In Fig. \ref{fig:6}, the corresponding plot with various $\alpha$ 
for $E_{Th}=\Delta_{0}$ is shown. 
$\rho(\varepsilon)$ is the normalized LDOS by its value in the 
normal state, where $\varepsilon$ denotes the quasiparticle energy measured from 
the Fermi surface. 
The curves $a$, $b$ and $c$ denote the LDOS at 
$x=-L/4$,  $-L/2$ and $-L$, respectively. 
Since DN is connected to the normal electrode at $x=-L$, 
$\rho(\varepsilon)=1$ is satisfied independent of $\varepsilon$ as 
shown in curves $c$ in Figs. \ref{fig:5} and \ref{fig:6}. 
As shown in Fig. \ref{fig:5}, 
the $\rho(\varepsilon)$ has a zero energy peak (ZEP) in DN 
(curves $a$ and $b$ in left and right panels). 
Even if $\alpha$ deviates from $0$, 
ZEP in LDOS does not vanish (curves $a$ and $b$ in the 
left and middle panels of Fig. \ref{fig:6}). 
This is because that 
quasiparticles with injection angle $\phi$ with 
$-\pi/2 + \alpha < \phi<\pi/2 -\alpha$ feel MARS and can contribute to the 
proximity effect. 
Exceptional case is $\alpha=\pi/2$, 
where  proximity effect is absent. 
Then the resulting $\rho(\varepsilon)=1$ independent of $\varepsilon$. 

\begin{figure}[tbh]
\begin{center}
\scalebox{0.4}{
\includegraphics[width=18.0cm,clip]{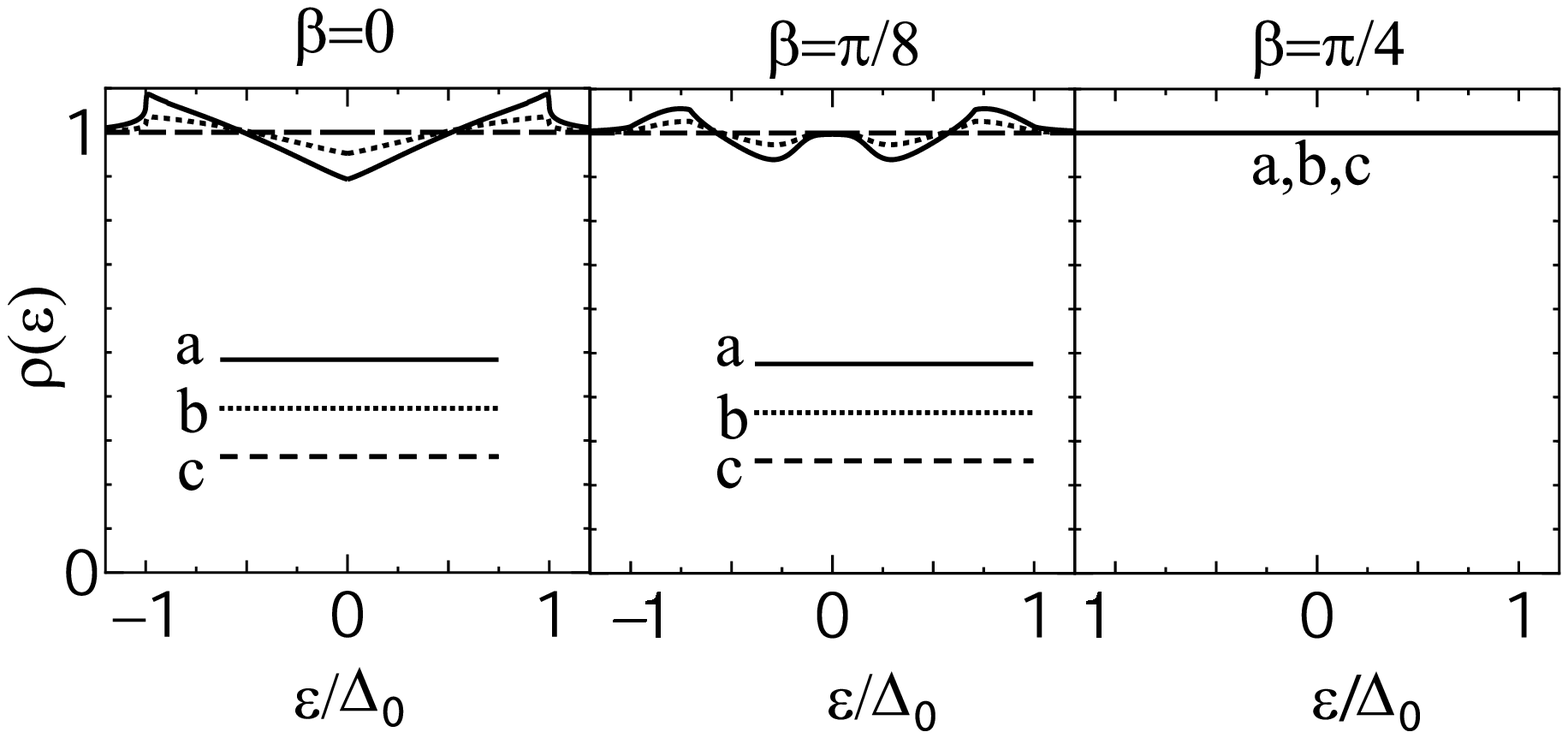}}
\end{center}
\caption{  Normalized  local density of states 
$\rho(\varepsilon)$ in DN for DN/USS junction with 
$Z=3$, $R_{d}/R_{b}=1$ and $E_{Th}=\Delta_{0}$. 
We have chosen $d$-wave superconductor 
with $\Delta_{\pm}=\Delta_{0}\cos[2(\theta \mp \beta)]$.  
$\beta=0$ (left panel), $\beta=\pi/8$ 
(middle panel), and $\beta=\pi/4$ (right panel). 
a, $x=-L/4$; b, $x= -L/2$; and c, $x=-L$. }
\label{fig:6i}
\end{figure}

In order to compare with DN/USS cases, we choose $d$-wave pair potential 
with $\Delta_{\pm}=\Delta_{0}\cos[2(\theta \mp \beta)]$, 
where $\beta$ denotes the angle between the normal to the interface and the 
crystal axis of $d$-wave superconductor \cite{PRB2004}. 
The corresponding $\rho(\varepsilon)$ is shown in Fig. \ref{fig:6i}. 
$\rho(\varepsilon)$ is nearly constant as a function of $\varepsilon$ and 
does not have a sharp ZEP  (Figs. \ref{fig:5} and \ref{fig:6}). 
In this case, although quasiparticles with injection angle 
$ \pi/4 -\beta< \mid \phi \mid < \pi/4 + \beta$ feel MARS, they 
can not contribute to the proximity effect. 
For $\beta=0$, MARS is absent and proximity effect becomes conventional one 
(see also Fig. 1 of \cite{PRB2004}). 
$\rho(\varepsilon)$ at $x=-L/4$ has a gap like structure (curve $a$ in the 
left panel of Fig. \ref{fig:6i}). 
Although $\rho(\varepsilon)$ at  $x=-L/4$ has a broad peak like structure 
for $\beta=\pi/8$, $\rho(0) \leq 1$ is satisfied contrary to the 
 DN/TS junction's case. 
For $\beta=\pi/4$, although $\sigma_{T}(eV)$ has a ZBCP, due to the absence of the proximity effect, $\rho(\varepsilon)=1$ for any case.  
Thus we can conclude that line shapes of $\rho(\varepsilon)$ in DN region
of DN/TS junctions are significantly different from 
those of DN/USS junctions.

\begin{figure}[tbh]
\begin{center}
\scalebox{0.4}{
\includegraphics[width=15.0cm,clip]{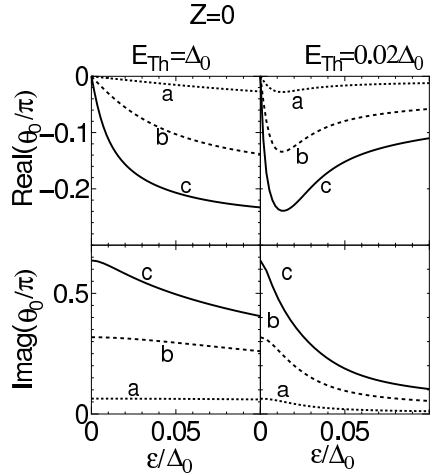}}
\end{center}
\caption{ Real(upper panels) and imaginary parts(lower panels) of $\protect%
\theta _{0}$ are plotted as a function of $\protect\varepsilon $ for 
$\alpha=0$ and $Z=0$ with $E_{Th}=\Delta_{0}$ (left panels) and 
$E_{Th}=0.02\Delta_{0}$ (right panels). 
a, $R_{d}/R_{b}=0.1$; b, $R_{d}/R_{b}=0.5$; and c, $%
R_{d}/R_{b}=1$. }
\label{fig:7}
\end{figure}

\begin{figure}[tbh]
\begin{center}
\scalebox{0.4}{
\includegraphics[width=15.0cm,clip]{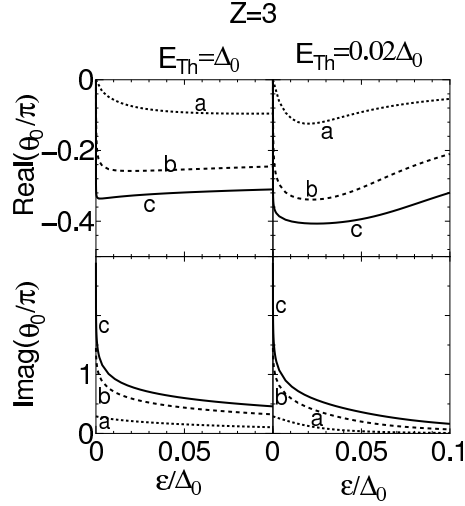}}
\end{center}
\caption{ Real(upper panels) and imaginary parts(lower panels) of $\protect%
\theta_{0}$ are plotted as a function of $\protect\varepsilon$ 
for $\alpha=0$ and $Z=3$. $E_{Th}=\Delta_{0}$ (left
panels) and $E_{Th}=0.02\Delta_{0}$ (right panels). 
a, $R_{d}/R_{b}=0.1$; b, $R_{d}/R_{b}=0.5$; and c, $R_{d}/R_{b}=1$.
}
\label{fig:8}
\end{figure}

\begin{figure}[tbh]
\begin{center}
\scalebox{0.4}{
\includegraphics[width=15.0cm,clip]{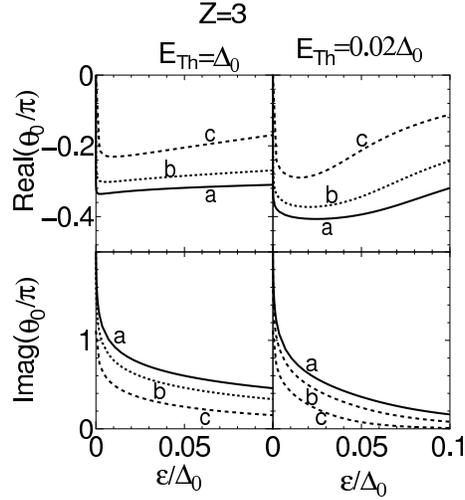}}
\end{center}
\caption{ Real(upper panels) and imaginary parts(lower panels) of $\protect%
\theta_{0}$ are plotted as a function of $\protect\varepsilon$ 
for $R_{d}/R_{b}=1$ and $Z=3$. $E_{Th}=\Delta_{0}$ (left
panels) and $E_{Th}=0.02\Delta_{0}$ (right panels). 
a, $\alpha=0$; b, $\alpha=0.25\pi$; and c, $\alpha=0.375\pi$.
}
\label{fig:8i}
\end{figure}

In Figs. \ref{fig:7} and \ref{fig:8}, 
the proximity parameter $\theta_{0}=\theta(x=0_{-})$ 
is plotted as a function of $\varepsilon$. 
In Fig. \ref{fig:8i}, $\alpha$ dependence of the 
$\theta_{0}$ is plotted. 
The magnitude of $\mathrm{Real(Imag)}(\theta _{0})$ is
increased with the increase of $R_{d}/R_{b}$. 
Contrary to the case of DN/USS junctions, 
$\mathrm{Real}(\theta _{0})$ vanishes at $\varepsilon=0$ 
while $\mathrm{Imag}(\theta _{0})$ remains nonzero at $\varepsilon=0$. 
For fully transparent case (Fig. \ref{fig:7}), 
$\mathrm{Real}(\theta _{0})$ and 
$\mathrm{Imag}(\theta _{0})$ 
decrease with the increase of $\varepsilon/\Delta_{0}$ for $E_{Th}=\Delta_{0}$. %
For $E_{Th}=0.02\Delta_{0}$, $\mathrm{Real}(\theta _{0})$ decreases 
from 0 and has a minimum at about $\varepsilon \simeq 0.02\Delta_{0}$. 
$\mathrm{Imag}(\theta _{0})$ decreases rather rapidly as 
compared to the case with $E_{Th}=\Delta_{0}$. 
For low transparent case with $Z=3$ (Fig. \ref{fig:8}), 
$\mathrm{Real}(\theta _{0})$ 
has a sudden change  around $\varepsilon \simeq0$. 
This sudden change is remarkable for 
the large magnitude of $R_{d}/R_{b}$ (see curves $b$ and $c$). 
With the increase of the magnitude of $\alpha$, 
the resulting magnitude of $\theta_{0}$ is reduced as shown in 
Fig. \ref{fig:8i}. 
For $\alpha=\pi/2$, the resulting $\theta_{0}$ is zero. \par
%

As seen from the sudden change of $\theta_{0}$ near zero energy 
for low transparent junction, we can imagine that 
there is a very small energy scale.
Actually, as seen from Fig. \ref{fig:1}, the width of 
$\sigma_{T}(eV)$ is significantly reduced for low transparent junction
with the increase of the magnitude of $R_{d}$.  
The sudden reduction of the magnitude of $\sigma_{T}(eV)$ around 
$eV \sim 0$ is related to the sudden change of $\theta_{0}$ 
around $\varepsilon=0$. 

\begin{figure}[tbh]
\begin{center}
\scalebox{0.4}{
\includegraphics[width=15.0cm,clip]{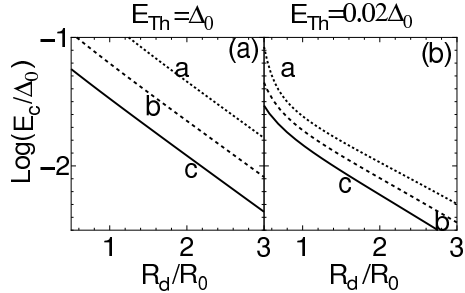}}
\end{center}
\caption{ $E_{C}$ is plotted as a function of $R_{d}/R_{b}$ 
for $E_{Th}=\Delta_{0}$ (left panel) and $E_{Th}=0.02\Delta_{0}$ 
(right panel). 
a, $Z=1$; b, $Z=2$; and c, $Z=3$.
}
\label{fig:9}
\end{figure}

\begin{figure}[tbh]
\begin{center}
\scalebox{0.4}{
\includegraphics[width=15.0cm,clip]{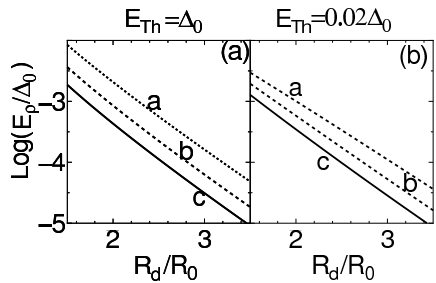}}
\end{center}
\caption{ $E_{\rho}$ is plotted as a function of $R_{d}/R_{b}$ 
for $E_{Th}=\Delta_{0}$ (left panel) and $E_{Th}=0.02\Delta_{0}$ 
(right panel). 
a, $Z=1$; b, $Z=2$; and c, $Z=3$.
}
\label{fig:10}
\end{figure}

In order to understand this small energy scale, 
we focus on the half width of the zero bias conductance peak of 
$\sigma_{T}(eV)$. 
We define the half width $E_{C}$  as 
$\sigma_{T}(E_{C})=\frac{1}{2}\sigma_{T}(0)$ and 
$\rho(E_{\rho})=\frac{1}{2}\rho(0)$, respectively.  
In Fig. \ref{fig:9}, 
$E_{C}$ is plotted as a function of $R_{d}/R_{0}$ for 
various $Z$. 
If the magnitude of $R_{d}/R_{0}$ is large, 
$E_{C}$ can be expressed by 
$E_{C}/\Delta_{0} \sim E_{C1}\exp(-C_{C}R_{d}/R_{0})$, 
where $E_{C1}$ and $C_{C}$ are independent of $R_{d}$. 
It is interesting that the magnitude of $C_{C}$ is almost 
independent of $Z$. 
As seen from left and right panels, 
the magnitude of $C_{C}$ is almost constant by the 
 change of $E_{Th}$. 
It is a very unique feature of the present proximity effect that 
$E_{C}$ has an exponential dependence of $R_{d}$ and the  
magnitude of $E_{C}$ is drastically reduced with the increase of $R_{d}$. 
The corresponding  
$\rho(\varepsilon)$ also has a sharp ZEP as shown in Fig. \ref{fig:5}.  
In order to understand this small energy scale, 
we define $E_{\rho}$  as $\rho(E_{\rho})=\frac{1}{2}\rho(0)$. 
In Fig. \ref{fig:10}, $E_{\rho}$ is plotted as a function of $R_{d}/R_{0}$. 
$E_{\rho}$ can be expressed by 
$E_{\rho}/\Delta_{0} \sim E_{\rho 1}\exp(-C_{\rho }R_{d}/R_{0})$ for 
the large magnitude of $R_{d}/R_{0}$  
where $E_{\rho 1}$ and $C_{\rho }$ are independent of $R_{d}$. 
Comparing the curvature of curves from $a$ to $c$, 
the magnitude of $C_{\rho}$ is almost independent of  
$Z$.  
As seen from left and right panels, 
the magnitude of $C_{\rho}$ is almost constant by the 
change of $E_{Th}$. 
The magnitude of $E_{\rho}$ is drastically suppressed with the increase of 
$R_{d}$ as in the case of $E_{C}$. 

\subsection{Properties at zero voltage}
\begin{figure}[tbh]
\begin{center}
\scalebox{0.7}{
\includegraphics[width=6.0cm,clip]{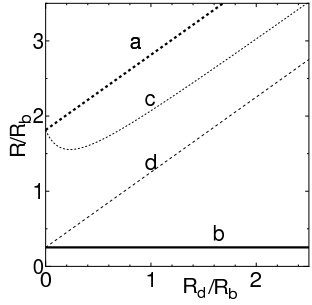}}
\end{center}
\caption{
Total zero voltage resistance of the junctions $R$ 
is plotted as a function of $R_{d}/R_{b}$ for $Z=1.5$ with  
a, $\alpha=0.25\pi$; and b, $\alpha=0.4\pi$.
The curve c  and d presents the same dependence 
for the DN/CSS junctions with   
$s$-wave superconductor 
and DN/USS junctions with $d_{xy}$-wave superconductor, 
respectively.}  
\label{fig:11}
\end{figure}

In this subsection, we focus on the zero voltage properties 
of DN/TS junctions with $p$-wave superconductor, where we can obtain
several analytical results without solving Usadel equation numerically. 
The $\theta_{0}$ is given by 
$\frac{ 2iR_{d} \cos \alpha}{R_{0}}$ by solving eq. (\ref{boundary1}).  
By using eqs. (\ref{ib0}) and (\ref{resistance}), 
the zero voltage resistance $R$ is given by
\begin{equation}
R=R_{0}
\{
\frac{{\rm tanh}\theta_{0i} }{I_{1}}
+
\frac{2}{[1 + \exp(2\mid \theta_{0i} \mid)]I_{1}+{\rm cosh}^{2}\theta_{0i} I_{2}}
\}
\label{exactresistance}
\end{equation}
\[
I_{1}=\int^{\pi/2-\alpha}_{-\pi/2 +\alpha}  \cos \phi d\phi
\]
\[
I_{2}=\int_{-\pi/2}^{-\pi/2 + \alpha}
\frac{2T^{2}(\phi) \cos\phi}{[2-T(\phi)]^{2}} d\phi
+ 
\int_{\pi/2-\alpha}^{\pi/2}
\frac{2T^{2}(\phi)\cos\phi}{[2-T(\phi)]^{2}} d\phi
\]
with $\theta_{0i}=2R_{d}\cos \alpha/R_{0} $.
First, we calculate 
the zero-voltage resistance ($\varepsilon \rightarrow 0)$
at different values of $\alpha$ as a function of $R_b/R_0$ 
for the DN/TS junctions with two extreme cases 
(curves $a$ and $b$ of Fig. \ref{fig:11}).
For $\alpha=\pi/2$, namely the $p_{y}$-wave case, 
$R$ increases linearly as a function of $R_{d}$, where no proximity effect 
appears (curve $a$ of Fig. \ref{fig:11}). 
For  $\alpha=0$, $p_{x}$-wave case, 
where $R=R_{0}/2$ is satisfied  independent of $R_{d}$ 
(curve $b$ of Fig. \ref{fig:11}). 
This anomalous $R$ dependence 
is a most striking feature by the enhanced proximity 
effect by the MARS. 
The corresponding result for the DN/CSS junctions with 
$s$-wave (curve $c$) and DN/USS junctions with 
$d_{xy}$-wave case (curve $d$) is also plotted as 
a reference. 
For $s$-wave case,  it is well known \cite{reflec} that 
$R$ has a reentrant behavior $\partial R/\partial R_{d}\mid_{R_{d}=0}<0$ 
as shown in curve $c$ of Fig. (\ref{fig:11}). 
In $p$-wave cases,  this reentrant behavior of $R$ 
does not appear. 
Actually, from the relation 
\begin{equation}
\left. \frac{\partial R}{\partial R_{D}} \right|_{R_{D}=0}
= 1 - \frac{ \cos^{2} \alpha}
{\left[\cos \alpha + \int^{\pi/2}_{\pi/2-\alpha} 
\frac{T^{2}(\phi)\cos \phi}{[2 -T(\phi)]^{2}} d\phi \right]^{2}},  
\end{equation}
we can show that $dR/dR_{D}\mid_{R_{D}=0}>0$ is always satisfied. 
For $d_{xy}$-wave case, due to the formation of the MARS
for all $\phi$ as in the case of 
$p_{x}$-wave junction, $R/R_{b}=R_{0}/(2R_{b})$ at $R_{d}=0$,  
which is identical to that for $p_{x}$-wave junction (curve $d$ of Fig. \ref{fig:11}). %
However, for nonzero $R_{d}$, 
$R/R_{b}=R_{0}/(2R_{b}) + R_{d}/R_{b}$ is satisfied, 
due to the absence of the proximity effect 
as discussed in our previous papers \cite{PRL2003,PRB2004}. 

\begin{figure}[htb]
\begin{center}
\scalebox{0.4}{
\includegraphics[width=15.0cm,clip]{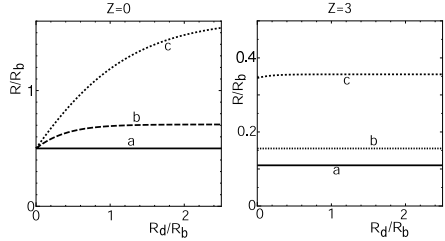}}
\end{center}
\caption{ 
Total zero voltage resistance of the junctions $R$ 
is plotted as a function of $R_{d}/R_{b}$ for 
$Z=0$ (left panel) and $Z=3$ (right panel).  
a, $\alpha=0$;  b, $\alpha=\pi/2$; 
and c, $\alpha=0.4\pi$. }
\label{fig:12}
\end{figure}

The $\alpha$ dependence of $R$ is shown in Fig. \ref{fig:12}. 
For $Z=0$, as shown in curves $b$ and $c$, $R$ increases gradually 
with $R_{d}/R_{b}$. However, for the large magnitude of $R_{d}/R_{b}$, 
the ratio of the increment is saturated and $R$ is almost constant as a function of $R_{d}/R_{b}$. 
For $Z=3$, $R$ is nearly constant with the increase of 
$R_{d}/R_{b}$. 
As seen from Eq. (\ref{exactresistance}), for sufficiently large magnitude of 
$R_{d}/R_{0}$, the second term in  Eq. (\ref{exactresistance}) becomes 
negligibly small  and 
$R$ converges to $R_{0}/(2\cos\alpha)$ independent of the 
magnitudes of  $Z$ and $R_{b}$. 
On the other hand, for small transparent junctions, $i.e.$,  
the large magnitudes of $Z$, 
we can neglect the term proportional to $I_{2}$
in Eq. (\ref{exactresistance}). 
In such a case, $R$ converges to be $R_{0}/(2\cos\alpha)$ 
independent of $R_{d}$. 
For $Z=3$ (see the right panel of Fig. \ref{fig:12}), 
$R$ saturates to be constant much more 
rapidly with the increase of $R_{d}/R_{b}$ as compared to the 
corresponding case for $Z=0$ (see the right panel of Fig. \ref{fig:12}).

Before closing this subsection, we pay attention to the 
zero voltage electrostatic potential distribution 
at zero temperature $\phi_{0}(x)$. 
Due to the enhanced proximity effect by the MARS, 
the spatial dependence of $\phi_{0}(x)$ is very unusual. 
The electrostatic potential in the DN region is expressed by 

\begin{equation}
\Phi(x)=\frac{1}{4}\int^{\infty}_{0} d\varepsilon 
{\rm Trace}[\hat{K}_{N}(x)]
=\int^{\infty}_{0} d\varepsilon f_{3N}(x) {\rm Real}[\cos \theta]. 
\end{equation}

\begin{figure}
\begin{center}
\scalebox{0.4}{
\includegraphics[width=16.0cm,clip]{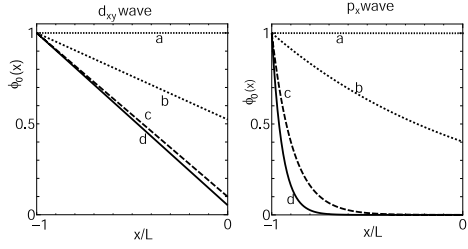}}
\end{center}
\caption{
Zero voltage electrostatic potential distribution 
at zero temperature $\phi_{0}(x)$ 
is plotted as a function of $x$ for $Z=3$. 
DN/USS junction with $d_{xy}$-wave superconductor (left panel) 
and  DN/TS junction with $p_{x}$-wave superconductor (left panel).  
a: $R_{d}/R_{b}=0$, 
b: $R_{d}/R_{b}=0.1$, 
c: $R_{d}/R_{b}=1$, 
and d: $R_{d}/R_{b}=2$.  
\label{fig:13}}
\end{figure}

The zero-bias voltage static potential distribution 
is defined  by \cite{Golubov}
\begin{equation}
\phi_{0}(x) =\lim_{V \rightarrow 0} \frac{\Phi(x)}{V}
\end{equation}
Since $f_{3N}(x)$ is expressed  by 
\begin{equation}
f_{3N}(x)=
\frac{ R_{b} + \frac{R_{d}<I_{b0}>}{L}\int^{0}_{x}\frac{dx}{\cosh^{2}
\theta_{imag}(x)} }
{R_{b} + \frac{R_{d}<I_{b0}>}{L}\int^{0}_{-L}\frac{dx}{\cosh^{2}
\theta_{imag}(x)} }f_{t0}, 
\end{equation}
with $\theta_{imag}(x)={\rm Imag}[\theta(x)]$, 
the resulting 
$\phi_{0}(x)$ is given by 
\begin{equation}
\phi_{0}(x)={\rm Real}[\cos \theta(x)]
\frac{ R_{b} + \frac{R_{d}<I_{b0}>}{L}\int^{0}_{x}\frac{dx}{\cosh^{2}
\theta_{imag}(x)} }
{R_{b} + \frac{R_{d}<I_{b0}>}{L}\int^{0}_{-L}\frac{dx}{\cosh^{2}
\theta_{imag}(x)} }
\end{equation} 
at zero temperature.  
In Fig. \ref{fig:13}, $x$ dependence of $\phi_{0}(x)$ is 
plotted both DN/USS junction with $d_{xy}$-wave superconductor and DN/TS junction with $p_{x}$-wave 
superconductor. 
Although quasiparticles always feel MARS both for 
$d_{xy}$-wave and $p_{x}$-wave cases, 
for $d_{xy}$-wave case the proximity effect is completely absent,
while for $p_{x}$-wave case, proximity effect is enhanced by MARS. 
For $d_{xy}$-wave case, since $\theta(x)=0$, $\phi_{0}(x)$ is given by
\begin{equation}
\phi_{0}(x)=\frac{LR_{0} -2xR_{d}}{L(R_{0} + 2R_{d})}. 
\end{equation}
Actually, as shown in Fig.  \ref{fig:13}, 
$\phi_{0}(x)$ has a linear $x$ dependence 
and decreases with $x$. 
This linear dependence corresponds to the ohm's rule. 
The magnitude of $\phi_{0}(x)$ is much reduced with the increase of 
$R_{d}/R_{b}$ (see curves $c$ and $d$).
For $p_{x}$-wave case, since $\theta(x)=2iR_{d}/R_{0}$ is satisfied, 
the resulting $\phi_{0}(x)$ is 
\begin{equation}
\phi_{0}(x)=\exp(-\frac{2(x+L)}{L}\frac{R_{d}}{R_{0}}). 
\end{equation}
Due to the enhanced proximity effect by MARS, $\phi_{0}(x)$ is reduced 
drastically with the increase of $R_{d}/R_{b}$ (see curves $c$ and $d$).

\subsection{Magnetic field dependence of the  conductance}
\begin{figure}[bh]
\begin{center}
\scalebox{0.4}{
\includegraphics[width=15.0cm,clip]{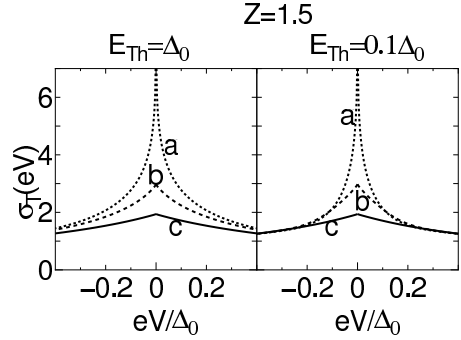}}
\end{center}
\caption{Normalized  conductance $\sigma_{T}(eV)$ for 
DN/TS junctions with $Z=1.5$ and $R_{d}/R_{b}=1$  where 
$p$-wave superconductor with $\alpha=0$ is chosen. 
We choose $E_{Th}=\Delta_{0}$ for the left panel and $E_{Th}=0.1\Delta_{0}$ 
for the right panel, respectively. 
a, $\hbar/\tau_{H}=0$; b, $\hbar/\tau_{H}=E_{Th}$; 
and c, $\hbar/\tau_{H}=10E_{Th}$. }
\label{fig:14}
\end{figure}

From the experimental view point, it is interesting to clarify 
how ZBCP in DN/TS junctions are influenced by the applied magnetic field $H$. 
It is known for conventional superconductor junctions, that 
coherence of electrons by the proximity effect 
is broken by the applied magnetic field or the magnetic impurity in DN 
\cite{Volkov,Yip} 
as shown in the following equation 
\begin{equation}
\hbar D\frac{\partial^{2}}{\partial x^{2}}\theta 
+ 2i \varepsilon \sin \theta
- \frac{\hbar}{\tau_{H}}\sin 2\theta = 0
\end{equation}
with $\hbar/\tau_{H}=6w^{2}D^{2}H^{2}$, 
where $w$ is the width of the DN region and  
$H$ is the applied magnetic field. 
With the increase of the magnetic field, the height of the 
ZBCP is suppressed as shown in Fig. \ref{fig:14}. 
The sharp peak structure around the zero voltage is significantly
suppressed when the magnitude of 
$\hbar/\tau_{H}$ is almost $10E_{Th}$. 
This is because the magnitude of ${\rm Im}\theta_{0}$ is significantly 
reduced by the increase of the magnitude of $\hbar/\tau_{H}$. 

\begin{figure}[bh]
\begin{center}
\scalebox{0.4}{
\includegraphics[width=15.0cm,clip]{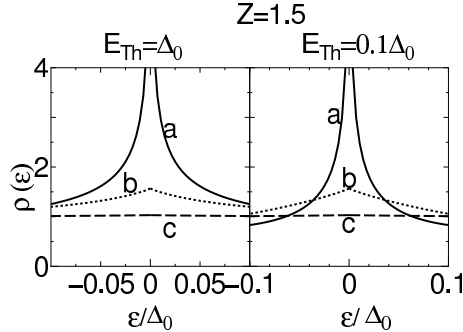}}
\end{center}
\caption{Normalized  local density of states of quasiparticle  $\rho(eV)$ for 
DN/TS junctions with $Z=1.5$ and $R_{d}/R_{b}=1$  where 
$p$-wave superconductor with $\alpha=0$ is chosen. We choose $x=-L/4$. 
$E_{Th}=\Delta_{0}$ for the left panel and $E_{Th}=0.1\Delta_{0}$ 
for the right panel, respectively. 
a, $\hbar/\tau_{H}=0$; b, $\hbar/\tau_{H}=E_{Th}$; 
and c, $\hbar/\tau_{H}=10E_{Th}$. }
\label{fig:14i}
\end{figure}

In order to understand the suppression of the proximity effect, 
we also look at $\rho(\varepsilon)$ in the DN region. 
The height of the ZEPs in $\rho(\varepsilon)$ 
both in the left $(E_{Th}=\Delta_{0})$ and right panels 
$(E_{Th}=0.1\Delta_{0})$
is reduced seriously with the increase of the magnitude of $\hbar/\tau_{H}$. 
For $\hbar/\tau_{H}=10E_{Th}$, $\rho(\varepsilon) \simeq 1$ for both cases 
(see curves $c$ in the left and right panels in Fig. \ref{fig:14i}). 
In this case, the proximity effect is almost absent and  
the ZBCP in $\sigma_{T}(eV)$ in curves $c$ in Fig. \ref{fig:14} originates 
purely from the formation of the MARS  at the DN/TS interface. 
The corresponding situation is realized in DN/USS junction with $d_{xy}$-wave 
superconductor (see Fig. \ref{fig:3}). 
Here, we estimate the threshold magnetic field which 
suppresses the sharp ZBCP or ZEP 
originating from the enhanced proximity effect. 
Here, $\hbar/\tau_{H}$ and 
$E_{Th}$ are given by $e^{2}w^{2}DH^{2}/(6\hbar)$ and $\hbar D/L^{2}$, 
respectively. 
We can define threshold magnetic field as  
$ 10E_{Th}= e^{2}w^{2}DH_{Th}^{2}/(6\hbar)$. 
We can easily show that $H_{Th} \simeq \frac{8\hbar}{e Lw}$. 
By choosing $L=1 \times 10^{-6}$[m] $w=1 \times 10^{-6}$[m], 
the resulting $H_{Th}$ is $5 \times 10^{-3}$ [Tesla]. 

\subsection{Conductance  for p+ip-wave superconductor}
\begin{figure}[bh]
\begin{center}
\scalebox{0.4}{
\includegraphics[width=15.0cm,clip]{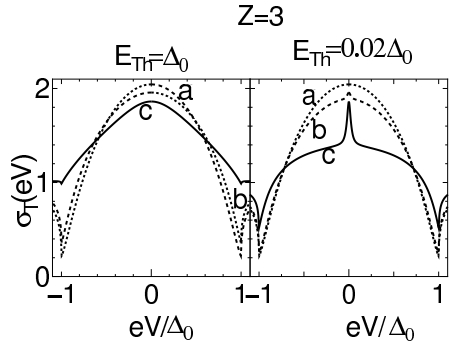}}
\end{center}
\caption{Normalized  conductance $\sigma_{T}(eV)$ for 
DN/TS junctions with $p_{x}+ip_{y}$ superconductor 
for $Z=3$ is plotted. We choose
$E_{Th}=\Delta_{0}$ for the left panel and $E_{Th}=0.02\Delta_{0}$ 
for the right panel, respectively. 
a, $R_{d}/R_{b}=0$; b, $R_{d}/R_{b}=0.1$; 
and c, $R_{d}/R_{b}=1$. }
\label{fig:15}
\end{figure}

In this subsection, we discuss the case where $p_{x}+ip_{y}$-wave 
superconductor is chosen as  TS(triplet superconductor). 
Stimulated by intensive studies about Sr$_{2}$RuO$_{4}$ after its discovery 
\cite{Maeno}, 
$p_{x}+ip_{y}$-wave superconductor, which is a 
$p$-wave superconductor with broken time symmetry, has 
a wide interest at the present. 
In this case, one of the important feature is that the 
MARS is formed only by the quasiparticle with 
perpendicular injection. 
The position of the resonance energy strongly depends on the 
injection angle of the quasiparticles. 
Also in this case, we can calculate the corresponding 
conductance $\sigma_{T}(eV)$ following similar calculations 
as in section 2. 
Since the  derivation of the matrix current and the 
relevant $<I_{b0}>$ is rather long, we will only show the 
final results. 
The details of the calculations will be  shown in elsewhere \cite{ppwave2005}. 
First, we look at the case with $R_{d}=0$, $i.e.$, ballistic junctions.
As shown in curves $a$ in the left and right panels of Fig. \ref{fig:15}, 
$\sigma_{T}(eV)$ has a rather broad peak \cite{tr1}. 
This is because that 
the position of the resonance energy deviates with the increase of 
the magnitude of the quasiparticle energy $\varepsilon$. 
For $E_{Th}=\Delta_{0}$, the magnitude of $\sigma_{T}(0)$ is 
suppressed with the increase of $R_{d}/R_{b}$, 
contrary to the case with DN/TS junctions without BTRSS 
as shown in Figs. \ref{fig:1}, \ref{fig:2}, and \ref{fig:4}. 
For $E_{Th}=0.02\Delta_{0}$, with the increase of $R_{d}/R_{b}$, 
$\sigma_{T}(0)$ has a sharp and narrow peak around zero voltage. 
This feature becomes prominent with the increase of $R_{d}/R_{b}$. 
%

\begin{figure}[bh]
\begin{center}
\scalebox{0.4}{
\includegraphics[width=15.0cm,clip]{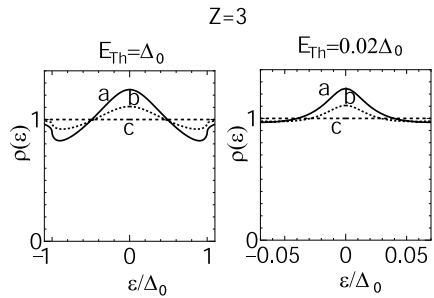}}
\end{center}
\caption{Normalized local density of states $\rho(\varepsilon)$ 
for DN/TS junctions with $p_{x}+ip_{y}$-wave  superconductor 
is plotted for $Z=3$. We choose 
$E_{Th}=\Delta_{0}$ for the left panel and $E_{Th}=0.02\Delta_{0}$ 
for the right panel, respectively. 
a, $x=-L/4$; b, $x=-L/2$; 
and c, $x=-L$. }
\label{fig:16}
\end{figure}

The corresponding plot for $\rho(\varepsilon)$ is shown in Fig. \ref{fig:16}.
$\rho(\varepsilon)$ is the normalized LDOS by its value in the 
normal state, where $\varepsilon$ denotes the quasiparticle energy measured from 
the Fermi surface. 
The curves $a$, $b$ and $c$ denote the LDOS at 
$x=-L/4$, $x=-L/2$ and $x=-L$, respectively.  
Since DN is connected to the normal electrode at $x=-L$, 
$\rho(\varepsilon)=1$ is satisfied independent of $\varepsilon$ as 
shown in curves $c$ in the left and right panels of 
Fig. \ref{fig:16}. 
$\rho(\varepsilon)$ has a zero energy peak (ZEP) in DN (curves $a$ and $b$). 
Although only the quasiparticles with perpendicular injection at the 
DN/TS interface feel the MARS, the ZEP in LDOS remains as in the case in Figs. \ref{fig:5} and 
\ref{fig:6}. 
The line shapes of LDOS for $p_{x}+ip_{y}$-wave case 
is significantly different from the corresponding ones for DN/USS junctions 
with $d$-wave superconductor (see Fig. \ref{fig:6i}). 
The existence of the ZEP in LDOS of DN is a remarkable feature peculiar to 
triplet superconductor. 
We think that this peculiar property which has never been expected in singlet 
junctions may serve as a guide to identify the triplet superconducting state. 

\section{Conclusions}

In the present paper, detailed theoretical investigation of the
voltage-dependent conductance of diffusive normal metal / insulator /
triplet superconductor (DN/TS) junctions is presented. 
We have provided the detailed derivation of
the formula of the matrix current presented in our previous paper \cite%
{pwave2004}. For the reader's convenience, we explicitly present
the retarded and the Keldysh parts of the matrix current 
in the DN/TS (diffusive normal metal / triplet superconductor ) junctions. 
Applying our general formula to DN/ insulator / $p$-wave 
superconductor junctions, we have obtained the following main results. 
In the present paper, by changing the barrier parameter at the interface 
$Z$, the resistance $R_{b}$ at the DN/TS interface, 
the resistance $R_{d}$ in DN, 
the Thouless energy $E_{Th}$ in DN and
the angle $\alpha$ 
between the normal to the interface and the lobe direction  of $p$%
-wave superconductor, we have studied the charge transport in 
DN/TS junctions in detail.  \par

\noindent 1. The zero bias conductance peak (ZBCP) 
is always seen in the line shape of $\sigma_{T}(eV)$ 
except for $\alpha = \pi/2$. 
The magnitude of $\sigma_{T}(0)$ is drastically enhanced 
with the increase of $R_{d}/R_{b}$. On the other hand,  $\sigma_{T}(eV)$ for 
diffusive normal metal / insulator / unconventional singlet superconductor 
(DN/USS) junctions with $d_{xy}$-wave superconductor shows a very different 
behavior. The magnitude of $\sigma_{T}(eV)$ is reduced with the increase of 
$R_{d}/R_{b}$ due to the absence of the proximity effect. 
For the large magnitude of $R_{d}/R_{0}$, 
the half width of the ZBCP is proportional to 
$\exp(-C_{C}R_{d}/R_{0})$, where $R_{0}$ is the Sharvin resistance 
at the DN/TS interface. 
$C_{C}$ is almost constant and independent of $Z$ and $E_{Th}$.  \par

\noindent 2. The LDOS has a ZEP except for the case with 
$\alpha = \pi/2$ where MARS is absent. 
The height of the ZEP is significantly 
enhanced  with the increase of $R_{d}/R_{0}$. 
For the large magnitude of $R_{d}/R_{0}$, 
the half width of the ZEP is proportional to 
$\exp(-C_{\rho}R_{d}/R_{0})$. 
$C_{\rho}$ is almost constant and independent of $Z$ and $E_{Th}$.  \par

\noindent 3. The proximity parameter at the DN/TS 
interface $\theta_{0}$ is $2i\cos\alpha R_{d}/R_{0}$  at $\varepsilon=0$ 
where  quasiparticle energy  $\varepsilon$  is measured from the Fermi energy. 
This unique feature  has never been expected for  
DN/USS or diffusive normal metal / insulator / conventional s-wave singlet superconductor (DN/CSS) case, 
where $\theta_{0}$ at $\varepsilon=0$ is always a real number.  \par

\noindent 4. 
The  total zero voltage resistance $R$ in the DN/TS junctions
is significantly  reduced by the enhanced proximity effect in the 
presence of the MARS. 
It is remarkable that when $R_{d}$ is sufficiently 
larger than the Sharvin resistance $R_{0}$, 
$R$ is reduced to be $R=R_{0}/(2 \cos \alpha)$,
which can become much smaller than the preexisting
lower limit value of $R$, $i.e.$,
$R_{0}/2 +R_{d}$.  
For low transparent junctions, 
$R$ is also reduced to be $R=R_{0}/(2 \cos \alpha)$ for any $R_{d}$. 
When all quasiparticles injected at the interface feel the
MARS, $R$ is reduced to be
$R=R_{0}/2$ irrespective of the  magnitude of $R_{d}$ and $R_{b}$.
This dramatic situation is realized for $\alpha=0$.  \par
\noindent 5. 
The sharp ZBCP or ZEP in LDOS 
due to the enhanced proximity effect is 
sensitive to the applied magnetic field $H$. 
The height of ZBCP or ZEP is significantly reduced by $H$. 
The threshold value of the magnetic field is 
$H_{Th} \simeq 8\hbar/(eS_{DN})$ where $S_{DN}$ denotes the 
magnitude of the area of DN region. \par

\noindent 6. 
We also choose $p_{x}+ip_{y}$-wave state as a model 
of chiral superconductor. Although only quasiparticles 
with perpendicular injection feel MARS, $\sigma_{T}(eV)$ 
has a ZBCP and LDOS has a ZEP,  due to the proximity effect. \par
\begin{figure}[bh]
\begin{center}
\scalebox{0.8}{
\includegraphics[width=15.0cm,clip]{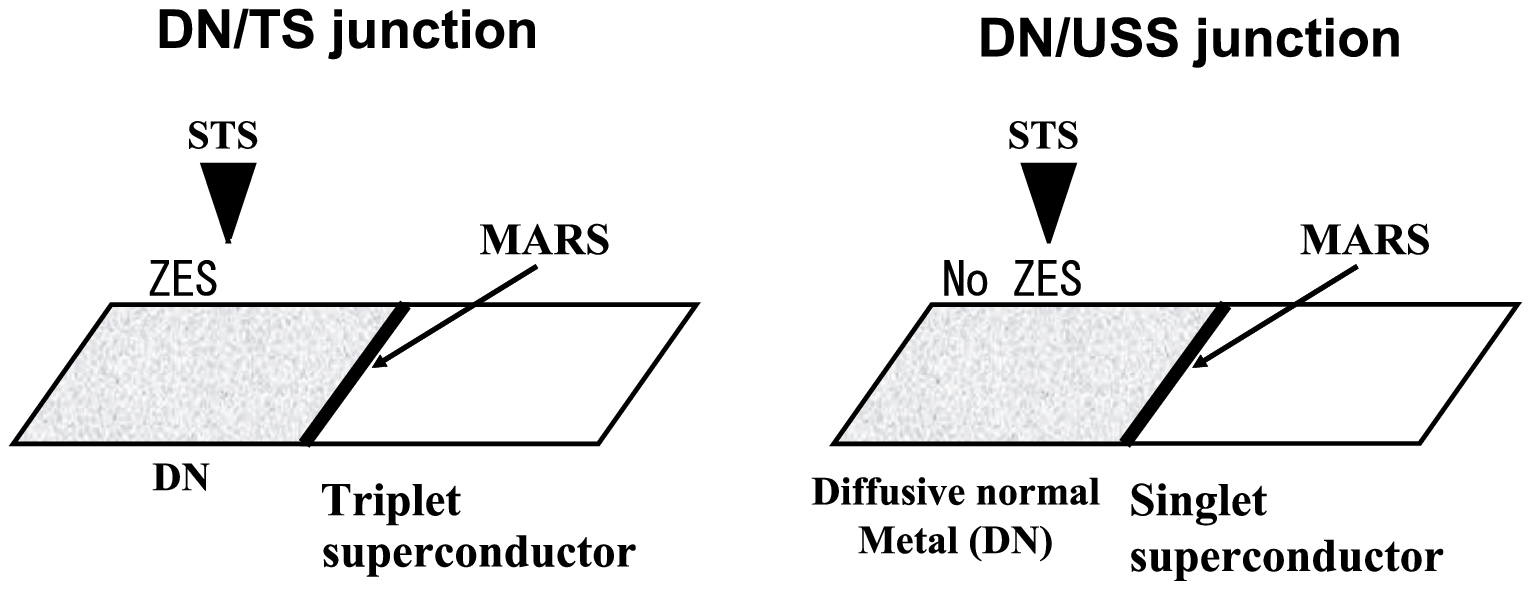}}
\end{center}
\caption{ Schematic illustration of the experimental  
setup to discriminate triplet superconducting state from 
singlet one. 
LDOS measured by STS only have a ZEP for DN/TS junctions.}
\label{fig:17}
\end{figure}
\begin{figure}[bh]
\begin{center}
\scalebox{0.8}{
\includegraphics[width=15.0cm,clip]{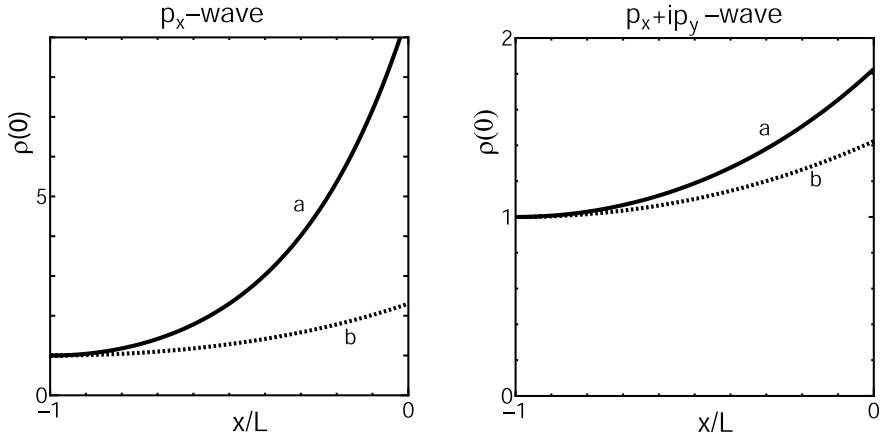}}
\end{center}
\caption{ Spatial dependence of the local density of state in DN region 
at zero energy is plotted for $p_{x}$-wave (left panel) 
and $p_{x}+ip_{y}$-wave (right panel) superconductors with $Z=1$.  
For $p_{x}$-wave case (left panel), $R_{d}/R_{b}=1$ (curve $a$), 
and  $R_{d}/R_{b}=0.5$ (curve $b$). 
For $p_{x}+ip_{y}$-wave case (left panel), 
$R_{d}/R_{b}=2$ (curve $a$), and  $R_{d}/R_{b}=1$ (curve $b$). Since we are concentrating on $\varepsilon=0$ case, $\rho(0)$ is independent of $E_{Th}$.  
}
\label{fig:21}
\end{figure}

These novel features have never been expected
either in DN/CSS or DN/USS junctions. Here, we propose an experimental 
setup which discriminate triplet superconducting state from singlet one 
by Scanning tunneling spectroscopy (STS) as shown in Fig. \ref{fig:17}. 
Only for triplet case, LDOS measured by STS has a ZEP. 
The spatial dependence of the height of ZEP can be 
observed by the proposed STS. 
According to our calculation discussed in the last section, 
we can obtain the LDOS for various $x$, 
where the positions of the  DN/TS interface is
denoted as $x=0$. 
As in our previous section, we assume that DN is attached to the 
normal electrode (N) at $x=-L$. 
The spatial dependence of $\rho(\varepsilon)$ 
for $\varepsilon=0$ is plotted in Fig. \ref{fig:21}  
both for $p_{x}$-wave  and $p_{x}+ip_{y}$-wave cases. 
Both for two cases, the magnitude of $\rho(0)$ decreases monotonically with the increase of the magnitude of the absolute value of 
$x/L$, where the position of 
the tip of STS changes from the DN/TS interface to the DN/N interface. 
We hope above spatial dependence of LDOS will be observed near future. 
As far as we know, the experiment of Knight shift measuring the 
uniform susceptibility is an only promising way to 
detect the triplet  superconducting state. 
Our proposal to identify the triplet superconducting state 
based on the proximity effect in the presence of the 
MARS is an innovational idea. \par
In the present paper, we have considered the normal metal / 
triplet or singlet superconductor junctions, where the 
mean free path in the normal metal region is much shorter than 
the length of it. 
Thus, it is possible to use Usadel equation.  
In the actual junctions, there is a possibility 
that the magnitude of the mean free path in normal metal becomes 
as the same order as its length.
To  describe this  situation, 
we have to  solve Eilenberger equation where the strength of the 
impurity scattering can be changed as an input parameter. 
Very recently, L\"{o}fwander \cite{Lofnew} calculated the 
local density of states in normal metal /$d$-wave superconductor junctions 
based on the Eilenberger equation. 
His obtained result is consistent with our theory when the impurity scattering 
effect becomes significant. 
It is an interesting problem to study the charge transport in 
triplet junctions based on the Eilenberger equation. 
\par
It is also possible to calculate tunneling conductance 
by the numerical simulation 
in the lattice model, where impurity potential 
is introduced in DN, where 
conductance is calculated for various samples which have different 
configuration of randomness each other \cite{recur1,recur2,Itoh,asano2004r}. 
Although it is difficult to take the large size of the system, 
the merit of the simulation is to take into account the 
impurity scattering effect  exactly. 
The prominent property, $ie.$,  
$R$ is completely independent of $R_{d}$ 
for $p_{x}$-wave junction, 
is verified by the numerical simulation \cite{asano2004r}. 
\par
A direct evidence of the mesoscopic interference effect 
by the proximity effect in unconventional superconductors 
has been reported in size dependence effect of high-$T_c$ superconductor 
junctions.
When the junction size becomes smaller, the conductance spectra are shown to be modified from the ballistic features. 
Since the high sensitivity of the ZBCP to the applied field was observed, 
the experimental data are consistent with the present results \cite{Hiromi}. 
Another way to detect the phase coherence in DN is to observe 
the LDOS of normal conductor in the vicinity of unconventional superconductors by tunneling spectroscopy.
There have been presented two different types of tunneling results on the LDOS observation of Au coated on YBCO.
Experimental results of Ref. \cite{SNS} show the presence of gap structure both for Au/YBCO(100) 
and Au/YBCO(110), while Asulin $et$ $al.$ observed ZBCP on Au/YBCO(110) when Au is thinner than 7nm, 
while the ZBCP suddenly disappears as the Au layer becomes thicker\cite{Sharoni}. 
These experimental reports are consistent with the present theory \cite{PRB2004} 
in the point that MARS can not penetrate into DN in the case of $d$-wave superconductors. 
However, since both of these measurements were performed 
on planer bilayer system (not mesoscopic size junctions), and the Au overlayer cannot be regarded as DN,
the detailed comparison 
of these results with the present theory seems to be inadequate.
On the other hand, 
it is a challenging issue to make a junction using Sr$_{2}$RuO$_{4}$
where ZBCP is already reported in tunneling spectroscopy \cite{e8}. 
In order to observe the enhanced proximity effect, we think the measurements 
in extremely small residual magnetic field environments are necessary.
This is because the ZBCP is easily suppressed even by the terrestrial magnetism.
\par
There are several problems which have not been discussed in the present paper. 
In the present study, we have focused on N/S junctions. The extension of
the circuit theory to long diffusive S/N/S junctions has been
performed by Bezuglyi \textit{et al.} \cite{Bezuglyi}. In S/N/S junctions,
the mechanism of multiple Andreev reflections produces the subharmonic gap
structures on I-V curves \cite{M1,M2,M3,M4,M5,M6,M7} and the situation
becomes much more complex as compared to N/S junctions. Moreover, in S/N/S
junctions with unconventional superconductors, MARS leads to the anomalous
current-phase relation and temperature dependence of the Josephson current
\cite{TKJ}. An interesting problem is an extension of the circuit theory to
S/N/S junctions with unconventional superconductors. 
Recent numerical simulation indicates the existence of the anomalous 
current phase relation in triplet junctions \cite{Asano2004J}. 
In the present paper, since we follow the quasiclassical Green's function
formalism, the impurity scattering is taken into account within the
self-consistent Born approximation. It is a challenging problem to study the
weak localization effects. 

%
The authors appreciate useful and fruitful discussions with 
A. Golubov, Y. Nazarov, Y. Asano, K.Kuroki, 
J. Inoue, H.Itoh, M. Kawamura, H. Yaguchi,  H. Takayanagi and Y. Maeno. 
This work was supported by the Core Research for Evolutional Science
and Technology (CREST) of the Japan Science and Technology Corporation
(JST). The computational aspect of this work has been performed at the
facilities of the Supercomputer Center, Institute for Solid State Physics,
University of Tokyo and the Computer Center. This work is supported by 
a Grant-in-Aid for the 21st Century  COE 
"Frontiers of Computational Science". 

%


\end{document}